\documentclass[twocolumn,tighten]{aastex61}
\usepackage{amssymb}
\usepackage{array,multirow}

\newcommand{\gyr}{{\rm{Gyr}}}

\newcommand{\msun}{{M_\odot}}

\newcommand{\pc}{{\rm{pc}}}
\newcommand{\henon}{{H\'enon}}
\newcommand{\metal}{{\rm{Z}}}
\newcommand{\kpc}{{\rm{kpc}}}
\newcommand{\rcobs}{{r_{c,\rm{obs}}}}
\newcommand{\rhl}{{r_{\rm{hl}}}}
\newcommand{\Sigmacobs}{{\Sigma_{c,\rm{obs}}}}
\newcommand{\popone}{{\tt{Pop1}}}
\newcommand{\poptwo}{{\tt{Pop2}}}
\newcommand{\Lto}{{L_{\rm{cut}}}}
\newcommand{\Deltarfifty}{{\Delta_{r50}}}
\newcommand{\Deltaa}{{\Delta_A}}
\newcommand{\nbh}{{N_{\rm{BH}}}}
\newcommand{\ncluster}{{N_{\rm{cluster}}}}
\newcommand{\rlim}{{r_{\rm{lim}}}}
\newcommand{\nrec}{{n_{\rm{rec}}}}
\newcommand{\ninject}{{n_{\rm{inj}}}}
\newcommand{\pmsigma}{\pm 1\sigma}

\shorttitle{Constraining $N_{\rm{BH}}$ in GCs}
\shortauthors{Weatherford, et al}

\begin{document}
\title{Predicting Stellar-Mass Black Hole Populations in Globular Clusters}

\author[0000-0002-9660-9085]{Newlin C. Weatherford}
\affiliation{Center for Interdisciplinary Exploration \& Research in Astrophysics (CIERA), Northwestern University, IL 60202, USA}
\email{newlinweatherford2017@u.northwestern.edu}

\author[0000-0002-3680-2684]{Sourav Chatterjee}
\affiliation{Tata Institute of Fundamental Research, Department of Astronomy and Astrophysics, Homi Bhaba Road, Navy Nagar, Colaba, Mumbai, 400005, India}
\affiliation{Center for Interdisciplinary Exploration \& Research in Astrophysics (CIERA), Northwestern University, IL 60202, USA}
\affiliation{Physics \& Astronomy, Northwestern University, IL 60202, USA}
\email{chatterjee.sourav2010@gmail.com}
\correspondingauthor{Sourav Chatterjee}

\author{Carl L. Rodriguez}
\affiliation{MIT-Kavli Institute for Astrophysics and Space Research, 77 Massachusetts Avenue, 37-664H, Cambridge, MA 02139, USA}
\affiliation{Center for Interdisciplinary Exploration \& Research in Astrophysics (CIERA), Northwestern University, IL 60202, USA}

\author[0000-0002-7132-418X]{Frederic A. Rasio}
\affiliation{Center for Interdisciplinary Exploration \& Research in Astrophysics (CIERA), Northwestern University, IL 60202, USA}
\affiliation{Physics \& Astronomy, Northwestern University, IL 60202, USA}

\begin{abstract}
Recent discoveries of black hole (BH) candidates in Galactic and extragalactic globular clusters (GCs) have ignited interest in understanding how BHs dynamically evolve in a GC and the number of BHs ($\nbh$) that may still be retained by today's GCs. Numerical models show that even if stellar-mass BHs are retained in today's GCs, they are typically in configurations that are not directly detectable. We show that a suitably defined measure of mass segregation ($\Delta$) between, e.g., giants and low-mass main-sequence stars, can be an effective probe to indirectly estimate $\nbh$ in a GC aided by calibrations from numerical models. Using numerical models including all relevant physics we first show that $\nbh$ is strongly anticorrelated with $\Delta$ between giant stars and low-mass main-sequence stars. We apply the distributions of $\Delta$ vs $\nbh$ obtained from models to three Milky Way GCs to predict the $\nbh$ retained by them at present. We calculate $\Delta$ using the publicly available ACS survey data for 47\ Tuc, M\ 10, and M\ 22, all with identified stellar-mass BH candidates. Using these measured $\Delta$ and distributions of $\Delta$ vs $\nbh$ from models as calibration we predict distributions for $\nbh$ expected to be retained in these GCs. 
For 47\ Tuc, M\ 10, and M\ 22 our predicted distributions peak at $\nbh\approx20$, $24$, and $50$, whereas, within the $2\sigma$ confidence level, $\nbh$ can be up to $\sim150$, $50$, and $200$, respectively.
\end{abstract}


\section{Introduction} \label{intro}
Recent discoveries of black hole (BH) candidates in Galactic and extragalactic globular clusters \citep[GCs;][]{2007Natur.445..183M,2013ApJ...777...69C,2012Natur.490...71S,2015MNRAS.453.3918M} have dramatically altered the traditional belief that GCs do not presently retain more than a few stellar-mass BHs \citep[e.g.,][]{1969ApJ...158L.139S}. Traditionally, it was believed that BHs, being significantly more massive than the average star, undergo rapid mass segregation to create a high-density and low-$N$ subcluster at the center of the host cluster. Frequent and energetic dynamical encounters in this BH-subcluster eject most BHs on a few $\gyr$ timescale \citep[e.g.,][]{1993Natur.364..421K,1993Natur.364..423S,2000ApJ...528L..17P}. Thus the GCs, being typically $\sim 12\,\gyr$ old, were expected to retain at most a couple of BHs at present. However, modern state-of-the-art simulations of massive star clusters indicate that BHs, unless ejected from the clusters due to natal kicks, actually have a much longer evaporation timescale than was previously believed \citep[e.g.,][]{2007MNRAS.379L..40M,2008MNRAS.386...65M,2013MNRAS.432.2779B}. Theoretically, this is because most BHs do not stay dynamically decoupled from the rest of the cluster for any prolonged period of time, as was assumed in past rate-based studies \citep[e.g.,][]{2013MNRAS.432.2779B,2013ApJ...763L..15M,2015ApJ...800....9M,2017ApJ...834...68C}. These simulations also suggest that the binary fraction in BHs typically remains low, both with BH and non-BH companions \citep[e.g.,][]{2015ApJ...800....9M,2017ApJ...834...68C}. In addition, due to 
the low duty cycles for the active state of a mass-transferring BH \citep[e.g.,][]{2004ApJ...601L.171K}, it was suggested that finding even a handful of mass-transferring BH candidates via X-ray and radio emissions likely indicates a much larger population of undetected retained BHs in these clusters 
\citep[e.g.,][]{2012Natur.490...46U}. 

The discovery of gravitational waves (GW) emitted from merging binary black holes (BBHs) 
by the LIGO/Virgo collaboration has reignited interest in understanding the astrophysical origins of BBHs \citep[e.g.,][]{2016PhRvL.116f1102A,2016ApJ...818L..22A,2016PhRvX...6d1015A,2016PhRvL.116x1103A,2017PhRvL.118v1101A}. It has been shown by several groups that high-mass and dense star clusters, such as the GCs, can be hotbeds for the dynamical production of BBHs that would merge in the local universe and be detected by LIGO, Virgo, and LISA \citep[e.g.,][]{2009ApJ...690.1370M,2010MNRAS.402..371B,2015PhRvL.115e1101R,2016PASA...33...36H,2016ApJ...816...65A,2016PhRvD..93h4029R,2016ApJ...824L...8R,2017ApJ...834...68C,2017ApJ...836L..26C,2017MNRAS.464L..36A}. Current rate analysis indicates that this dynamical formation channel may account for at least half of all BBH mergers LIGO will detect \citep[e.g.,][]{2016PhRvD..93h4029R} simply from GCs that survive to present day. Indeed, the dynamical assembly process for BBHs is so efficient that even young massive star clusters can significantly contribute and might even double the estimated overall merger rates from the dynamical formation channel \citep[e.g.,][]{2014MNRAS.441.3703Z,2017MNRAS.467..524B,2018MNRAS.473..909B,2018arXiv180506466B}. The retention fraction of BHs as a function of time not only affects the dynamical formation rate of BBHs, the energetic encounters involving BHs also can dramatically alter the overall evolution, and even long-term survival, of the host clusters \citep[e.g.,][]{2004ApJ...608L..25M,2007MNRAS.379...93H,2007MNRAS.379L..40M,2008MNRAS.386...65M,2017ApJ...834...68C}. Thus the long-term retention and dynamical evolution of BHs in massive and dense star clusters is of great interest for several branches of astrophysics. 

While it is beyond doubt that massive star clusters form a large number of BHs, $\sim10^{-3}N$, where $N$ is the initial number of stars, e.g., assuming the initial stellar mass function (IMF) given by \citealt{2001MNRAS.322..231K}, how long they remain bound to the clusters and take part in the internal dynamics depends on several assumptions that lack strong observational constraints, including major ones such as the distribution of natal kicks the BHs receive and the mass function of BHs at birth \citep[e.g.,][]{2014MNRAS.439.2459H}. GCs in the Milky Way (MW) provide a unique laboratory to test the long-term retention of BHs in high-mass, high-density, and old star clusters, shown to be ideal for the creation of high-mass BBHs merging in the local universe. 

Observationally inferring the existence of a large population of retained BHs in a GC, however, is not straightforward. Although it was initially suggested that a GC hosting a BH candidate likely hosts a large number of undetected BHs \citep[e.g.,][]{2012Natur.490...46U}, results from modern simulations suggest that the number of mass-transferring BHs in a GC at any given time is not correlated with the total number of BHs ($\nbh$) retained in that GC at the time \citep{2017ApJ...836L..26C,2018ApJ...852...29K}. Hence, contrary to the initial expectations, detection of a mass-transferring BH candidate 
does not necessarily indicate the existence of a large population of undetected BHs in that 
cluster. Interestingly, the GCs that host the discovered BH candidates also do not show any apparent 
trends or specialty in their easily observable global properties such as the 
core radius, concentration, and mass \citep[e.g.,][]{2013ApJ...777...69C,2012Natur.490...71S,2015MNRAS.453.3918M}. 

Simulations also show that it is hard to infer the existence of a large population of BHs simply from the observed structural profile of a GC since the BH-mediated dynamics typically leave 
little signature in the observed light profile of the overall cluster, apart from making the cluster
appear puffier \citep[e.g.,][]{2017ApJ...836L..26C}. Indeed, several groups have suggested that detailed analysis of structural 
features, such as large core radius and low central density, may be indicative of the presence of a large population of 
retained BHs in a GC \citep[e.g.,][]{2004ApJ...608L..25M,2007MNRAS.379...93H,2015ApJ...800....9M,2017ApJ...834...68C,2017MNRAS.464L..36A,2018arXiv180100795A,2018arXiv180205284A}. 
However, there is always the ambiguity whether a GC is currently puffy because of BH dynamics-mediated energy production or simply because it was born puffy (equivalently, with a long initial relaxation time). It was also suggested that the radial variation in the present-day slope of the stellar mass function may contain signatures of the cluster's dynamical history, even revealing, to some degree, the presence of a dynamically relevant BH population \citep[e.g.,][]{2016MNRAS.463.2383W,2017MNRAS.471.3845W}. However, measuring the mass function of a GC in many radial bins is challenging and requires consolidating observations from different space- and ground-based instruments to obtain enough radial coverage of a real GC. 

BHs affect the overall evolution of a GC primarily through mass-segregation. Via gravitational encounters lighter stars typically gain energy and heavier stars lose energy. As a result, the heavier stars sink into the gravitational potential of a cluster on a timescale proportional to the product of the local relaxation time and the ratio of the average mass of nearby stellar species to the mass of the heavier stars \citep[e.g.,][]{1984MNRAS.210..763L,1995ApJ...452L..33K,2003gmbp.book.....H,2004ApJ...604..632G}. GCs typically are older than their relaxation times, so they are expected to be mass segregated. The resultant mass segregation is driven by the most massive species in a cluster at any given time. Hence, while a large number of BHs are still present, they dominate the central part of a GC and drive lower-mass stellar species outward. Though the BHs cannot be observed directly, bright stars from different mass ranges are observable and their relative locations can be compared. Since the degree of mass segregation is directly related to the internal dynamical evolution of all stellar species, bright or dark, several groups have proposed that this can be used to infer the existence of hard-to-observe dark remnants in a GC. In particular, a dynamically significant dark population 
at the center of a GC quenches the level of mass segregation ($\Delta$) between bright stars of different mass ranges \citep[e.g.,][]{2008MNRAS.386...65M}. For this reason, using $\Delta$ to infer the existence of an intermediate-mass BH (IMBH) at the center of a GC was proposed more than a decade ago \citep[e.g.,][]{2004ApJ...613.1143B,2007MNRAS.374..857T}. More recently, this measure has been used to put upper limits on the mass of an IMBH in real clusters by \citealt{2016ApJ...823..135P}. Using timescale arguments and somewhat idealized treatments of BH retention, \citet{2016MNRAS.462.2333P} suggested that the lack of mass-segregation between the blue straggler stars (BSs) and stars near the main-sequence turn-off (MSTO) in NGC 6101 may be due to an undetected population of retained BHs. \citet{2016ApJ...833..252A} also showed that 
the number of retained dark remnants can quench mass segregation between BSs and a set of lower-mass reference stars.  

Our goal in this study is to construct an `observer-friendly' mass segregation parameter that can be easily used independent of the GC. In general, to quantify mass segregation, one must define two 
stellar populations different in their typical masses and construct a measure of the difference in their radial locations. Using BSs to construct the heavier visible population, similar to some of the above-mentioned studies, has some shortcomings. For example, BSs are typically low in number. Especially, production of BSs is expected to be quenched due to the presence of a large BH population in a cluster \citep{1993Natur.364..423S}. Production efficiency of BSs also varies among specific clusters depending on several initial cluster properties (e.g., binary fraction, central density) and 
internal dynamics \citep[e.g., recently][]{2007ApJ...661..210L, 2013ApJ...777..106C, 2013ApJ...777..105S,2018arXiv180707679L}. Moreover, today's BSs, by definition, had their masses changed at some unconstrained earlier epoch \citep[e.g.,][]{1995ApJ...445L.117L,1996ApJ...468..797L,2002ApJ...568..939L,1997ApJ...487..290S,2001ApJ...548..323S,2009MNRAS.395.1822C}, which makes it harder to understand their mass segregation. Observationally, identifying BSs requires complicated and sometimes ad-hoc cuts on the color-magnitude diagram (CMD) primarily 
dependent on the the width of the single and binary main-sequences. Even the shape of the 
CMD and thus the separation of the BSs from the main-sequence is filter-dependent \citep[e.g.,][]{2013ApJ...770...45D}. 

In order to avoid all of the above difficulties, we simply define the two populations as follows. We use the giant stars and near-turn-off main sequence stars to represent the heavier population. We use low-mass main-sequence stars populating a 
particular region in the CMD defined relative to the magnitude of the MSTO to constitute the lighter population. Since both populations are anchored to 
the MSTO in the CMD, the population definitions are simpler to implement consistently and can easily be adjusted for specific clusters. Additional advantages are that the number of stars in both sets are large and that the majority of the members in both sets are simple, undisturbed stars -- i.e., their stellar properties are not affected by the internal dynamics of the cluster, but their radial 
locations are affected by the overall dynamics. 

We use a large grid of state-of-the-art numerical models to relate $\Delta$ calculated using the two populations with different average stellar masses to $\nbh$ retained in these models. We show that $\Delta$ indeed can be used to infer $\nbh$ retained in these cluster models. Calibrations connecting $\Delta$ with $\nbh$ obtained from these models are then used to infer $\nbh$ in a handful of real GCs in the Milky Way (MW) using the publicly available ACS survey data \citep{2007AJ....133.1658S}. 

We describe our models and define the stellar populations used to quantify mass segregation in \autoref{S:models}. We show our key results connecting $\Delta$ with $\nbh$ in our models in \autoref{S:model_results}. In \autoref{S:observational_results}, we show how the dependence between $\Delta$ and $\nbh$ can be used for real GCs to infer the $\nbh$ retained by them by selecting three GCs in the MW containing known BH candidates, and for which we can calculate $\Delta$ using the publicly available ACS survey data \citep{2007AJ....133.1658S}. We use the theoretically calibrated relations connecting $\nbh$ and $\Delta$ and predict the expected $\nbh$ in these GCs based on the measured $\Delta$ in \autoref{S:nbh_predictions}. Finally, we summarize our results and conclude in \autoref{S:summary}.

%%%%%%%%%%%%%%%%%%%%%%%%%%%%%%%%%%%%%%%%%%%%%%%%%%%%%%%%%%%%%%%%%%%%%%%%%%%%%%%%%%%%%%%%%%%%%%%%%
%
\section{Numerical Models and Extraction of Model Parameters} \label{S:models}
%
%%%%%%%%%%%%%%%%%%%%%%%%%%%%%%%%%%%%%%%%%%%%%%%%%%%%%%%%%%%%%%%%%%%%%%%%%%%%%%%%%%%%%%%%%%%%%%%%%
We create a large grid of cluster models using our \henon-type \citep{1971Ap&SS..13..284H, 1971Ap&SS..14..151H} Cluster Monte Carlo code (CMC) that has been developed and rigorously tested over the last 17 years \citep{2000ApJ...540..969J,2001ApJ...550..691J, 2003ApJ...593..772F, 2007ApJ...658.1047F, 2010ApJ...719..915C, 2012ApJ...750...31U,2013MNRAS.429.2881C}. For recent updates and validation of CMC see \citep{2013ApJS..204...15P, 2015ApJ...800....9M, 2016MNRAS.463.2109R}.

We vary the initial model properties motivated by observational constraints from high-mass young star clusters, thought to be similar in properties for GC progenitors \citep[e.g.,][]{2007A&A...469..925S,2010ApJ...719..915C,2014MNRAS.439.2459H}. The initial $N$ is between $2\times10^5$ and $2\times10^6$, the metallicity ($\metal$) is between $\metal/\metal_\odot=0.005$ and $1$, virial radius is between $r_v/\pc=1$ and $2$, galactocentric distance is between $R_g/\kpc=1$ and $20$, and the initial binary fraction is between $f_b=0.04$ and $0.1$. The positions and velocities of single stars and center of mass of binaries are drawn from a King profile with concentration $w_0=5$ \citep{1966AJ.....71...64K}. Stellar masses (primary mass in case of a binary) are drawn from the IMF for star clusters given in \citet{2001MNRAS.322..231K} between $0.08$ and $150\,\msun$. Binaries are assigned by randomly choosing $N\times f_b$ stars independent of radial position and mass and assigning a secondary adopting a uniform mass ratio ($q$) between $0.08/m_p$ and $1$, where $m_p$ denotes the primary mass. Binary 
periods are flat in log intervals and eccentricities are thermal. We include all relevant physical processes, such as two-body relaxation, strong binary-mediated scattering, and galactic tides. 

Single and binary stellar evolution is followed using SSE and BSE packages \citep{2000MNRAS.315..543H, 2002MNRAS.329..897H} updated to include our latest understanding on stellar winds \citep[e.g.,][]{2001A&A...369..574V, 2010ApJ...714.1217B} and BH formation physics \citep[e.g.,][]{2002ApJ...572..407B,2012ApJ...749...91F}. Neutron stars (NS) are given natal kicks drawn from a Maxwellian with $\sigma=265\,{\rm{kms^{-1}}}$. The maximum NS mass is fixed at $3\,\msun$. The BH mass spectrum (any remnant above the maximum NS mass is considered a BH) depend on the metallicities and pre-collapse mass \citep{2012ApJ...749...91F}. Natal kicks for BHs are given based on results from \citet{2002ApJ...572..407B,2012ApJ...749...91F}. Namely, a velocity is first drawn from a Maxwellian with $\sigma=265\,{\rm{kms^{-1}}}$ and is then scaled down based on the metallicity-dependent fallback of mass ejected due to supernova. These prescriptions lead to $\sim10^{-3}N$ retention of BHs immediately after they form. More detailed descriptions and justifications are given in past work \citep[e.g.,][]{2015ApJ...800....9M,2016MNRAS.458.1450W,2017MNRAS.464L..36A}. However, note that the primary results in this work do not depend on the exact prescriptions for BH natal kicks, provided that a dynamically significant BH population remains in the cluster post-supernova. These results are also expected to depend indirectly on the BH birth mass function, via modest differences it may create in the average stellar mass of the cluster at late times. 

Finally, to ensure that our results are not affected by clusters close to dissolution -- at that point, the assumption of spherical symmetry in CMC is incorrect -- only models that reached $12\, \gyr$ while retaining $\geq30\%$ of their initial $N$ were included in the analysis. In total, we use $37$ different combinations of initial properties and a total of $153$ separate models including multiple realizations of models using the same combination of initial properties. Detailed descriptions of each model are given in \autoref{T:modellist}.

\begin{deluxetable*}{|ccc|ccccc|}
\tabletypesize{\scriptsize}
\tablecolumns{8}
\tablewidth{0pt}
\tablecaption{Initial Properties of Cluster Models}
\tablehead{
%    \multicolumn{3}{|c|}{\textbf{Counts}} & \multicolumn{5}{c|}{\textbf{Initial Conditions}} \\
    Model & Runs & Snapshots & $r_v/\pc$ & $\metal/\metal_\odot$ & $R_g/\kpc$ &
    $N$ & % \ \left(\times10^6\right)
    $f_b$}
\startdata
 1  &  9 & 36 & 1 & 0.01  & 20 & 2e5 & 10 \\
 2  &  5 & 20 & 1 & 0.01  & 20 & 5e5 & 10 \\ 
 3  &  2 &  8 & 1 & 0.01  & 20 & 1e6 & 10 \\
 4  &  2 &  8 & 1 & 0.01  & 20 & 2e6 & 10 \\ \hline
 5  & 12 & 48 & 1 & 0.05  &  8 & 2e5 & 10 \\
 6  &  5 & 20 & 1 & 0.05  &  8 & 5e5 & 10 \\
 7  &  6 & 20 & 1 & 0.05  &  8 & 8e5 & 10 \\ 
 8  &  2 &  8 & 1 & 0.05  &  8 & 1e6 & 10 \\
 9  &  2 &  8 & 1 & 0.05  &  8 & 2e6 & 10 \\\hline
10  &  5 & 20 & 1 & 0.25  &  2 & 5e5 & 10 \\ 
11  &  2 &  8 & 1 & 0.25  &  2 & 1e6 & 10 \\
12  &  2 &  8 & 1 & 0.25  &  2 & 2e6 & 10 \\ \hline %\hline
13* &  4 & 12 & 2 & 0.005 &  8 & 8e5 & 10 \\ \hline
14  & 10 & 36 & 2 & 0.01  & 20 & 2e5 & 10 \\
15  &  5 & 20 & 2 & 0.01  & 20 & 5e5 & 10 \\
16  &  2 &  8 & 2 & 0.01  & 20 & 1e6 & 10 \\
17  &  2 &  8 & 2 & 0.01  & 20 & 2e6 & 10 \\ \hline 
18  &  1 &  4 & 2 & 0.015 & 20 & 8e5 &  5 \\ \hline
19  &  8 & 26 & 2 & 0.025 &  8 & 8e5 & 10 \\ \hline
20  &  6 & 14 & 2 & 0.05  &  8 & 2e5 & 10 \\
21  & 13 & 50 & 2 & 0.05  &  8 & 2e5 & 10 \\
22  &  5 & 20 & 2 & 0.05  &  8 & 5e5 & 10 \\
23  &  2 &  8 & 2 & 0.05  &  8 & 1e6 & 10 \\
24  &  2 &  8 & 2 & 0.05  &  8 & 2e6 & 10 \\
25  &  1 &  4 & 2 & 0.05  &  8 & 8e5 &  4 \\
26  &  2 &  8 & 2 & 0.05  &  8 & 8e5 &  5 \\
27  &  5 & 17 & 2 & 0.05  &  8 & 8e5 & 10 \\ \hline
28  &  1 &  4 & 2 & 0.15  &  4 & 8e5 &  5 \\ \hline
29  &  3 & 12 & 2 & 0.25  &  2 & 5e5 & 10 \\
30  &  2 &  8 & 2 & 0.25  &  2 & 1e6 & 10 \\
31  &  2 &  8 & 2 & 0.25  &  2 & 2e6 & 10 \\
32  &  5 & 16 & 2 & 0.25  &  8 & 8e5 & 10 \\ \hline
33  &  1 &  4 & 2 & 0.35  &  2 & 8e5 &  5 \\ \hline
34* &  8 & 21 & 2 & 0.5   &  8 & 8e5 & 10 \\ \hline
35* &  4 & 11 & 2 & 0.75  &  8 & 8e5 & 10 \\ \hline
36* &  4 & 11 & 2 & 1.0   &  8 & 8e5 & 10 \\
37* &  1 &  4 & 2 & 1.0   &  1 & 8e5 &  5 \\
\enddata
\tablecomments{`Model' denotes cluster models with a particular set of initial properties. `Runs' denote the number of realizations performed. `Snapshots' denote the number of snapshots taken into our analysis between $9$ and $12\,\gyr$. $r_v$, $\metal$, $R_g$, $N$, and $f_b$ denote the initial virial radius, stellar metallicity, galactocentric distance, number of objects, and binary fraction, respectively. `*' denotes that these models were not used for $\Delta$--$\nbh$ calibration because the $\metal$ of these models are not representative of the three GCs of interest in this study (\autoref{S:observational_results}). }
\label{T:modellist}
\end{deluxetable*}

\subsection{``Observing" Model Clusters} \label{S:observing_cluster_models}
CMC periodically generates the dynamical and stellar properties of all single and binary stars including the luminosity ($L$), temperature ($T$), and radial positions. We assume spherical symmetry and project the radial positions of all single and binary stars in two-dimensions to create sky-projected snapshots of the cluster models at different times. Taking into account the typical age range of MWGCs, we use up to $4$ snapshots for each model realization, corresponding roughly to $t=9$, $10$, $11$, and $12\, \gyr$. In some model realizations, CMC did not output a snapshot near enough to all $4$ of these standardized ages (within $0.25\, \gyr$). Such realizations therefore contributed fewer than $4$ snapshots to the analysis. The number of snapshots used in our analysis is included for each model in \autoref{T:modellist} for reference. In total, we use $554$ snapshots.
For each single star we calculate the $T$ from $L$ and the stellar radius $R$ (given by BSE) assuming a black-body. All binaries are treated as unresolved sources. In the case of a binary, we use the total luminosity $L=L_1+L_2$ and a $L$-weighted temperature given by  

\begin{equation}
T_{\rm{eff}} = \frac{T_1 L_1 + T_2 L_2}{L_1 + L_2}
\end{equation}

We account for statistical fluctuations by performing $25$ realizations of 2D projections for each snapshot selected as above by varying the random seed. For each realization of the 2D-projected snapshots we calculate the core radius ($\rcobs$) and the central surface luminosity density ($\Sigmacobs$) by fitting an analytic approximation of the King model \citep[Eq. 18;][]{1962AJ.....67..471K} to the cumulative luminosity profile \citep[e.g.,][]{2017ApJ...834...68C}. We also calculate the half-light radius ($\rhl$) as the sky-projected distance from the center within which half of the total cluster light is contained.  

\begin{figure}
\plotone{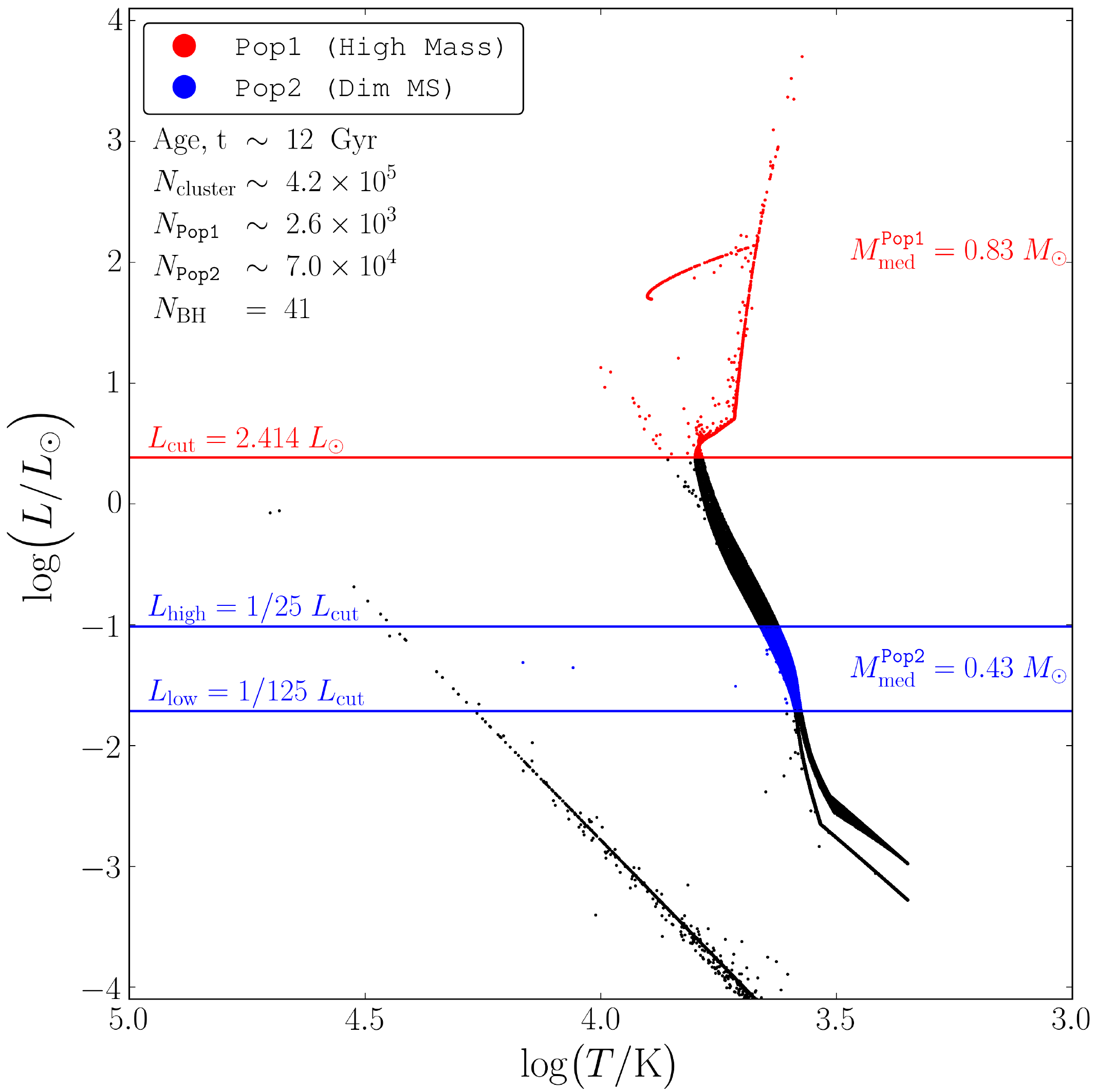}
\caption{Example Hertzsprung-Russel Diagram (HRD) from a model at $t=12\,\gyr$ (model no. 6; \autoref{T:modellist}) showing the two stellar populations compared for measuring the level of mass segregation. Each dot represents a single or binary star (all binaries are considered unresolved). Red and blue denote our high-mass (\popone) and low-mass (\poptwo) populations, respectively. Each population is defined with respect to the luminosity at the point of highest temperature on the MS (boundaries are delineated). The median masses for \popone\ and \poptwo\ are shown in the figure. Relevant cluster properties at the time of this snapshot such as $N$, $\nbh$, and the respective numbers in each stellar population ($N_\popone$ and $N_\poptwo$) are also included. Defining stellar populations this way ensures a high number of stars in each population. 
}
\label{f:1}
\end{figure}

%
%\newpage
\subsection{Population Selection} \label{S:population_selection}
In general, quantifying $\Delta$ in a GC requires comparison between the radial distributions of multiple stellar populations sufficiently different in their average masses \citep[e.g.,][]{2013ApJ...778...57G}. However, mass is not directly measured in real clusters -- stellar luminosities are, and can be used as a proxy for mass, especially in the MS \citep[e.g.,][]{1994sipp.book.....H}. 
While any two populations sufficiently different in their average masses could work for our study, devising a recipe that can be applied consistently to numerical models as well as most real GCs is somewhat challenging and to some degree a matter of choice. To ascertain maximum signal strength the two populations must have characteristic masses as different as possible, but at the same time, the lighter population must not be so faint that they are hard to observe in a real GC. In addition, both populations must contain large enough numbers of stars to limit statistical scatter. 
We decide to anchor the definitions of the two populations to the location of the MSTO, the most prominent feature on any CMD.
While it is easy to uniquely define the MSTO in theoretical models simply by finding the $L$ above which no H-burning single stars exist, this definition is not usable for observed clusters. Instead, to be consistent between the theoretical models and observed clusters, we define MSTO at $L=\Lto$ where the MS stars (excluding blue stragglers) exhibit the highest temperature. The high-mass population includes all stars with $L>\Lto$, consisting mostly of giants with some blue stragglers and high-mass MS stars near the turnoff, this heavier population will henceforth be referred to as \popone\ (\autoref{f:1}).
The lighter population, \poptwo, is selected to be the MS stars with $\Lto/125\leq L\leq \Lto/25$. 
This definition allows us to unambiguously and consistently define the two populations from the CMDs of observed GCs without the need for model-dependent methods such as isochrone fitting in order to find the location of the MSTO (\autoref{S:popselec2}). 

The limits in $L$ for \poptwo\ are chosen to optimize the signal strength for mass segregation while making sure that the equivalent area on a CMD for real GCs is available from the ACS GC survey \citep{2007AJ....133.1658S} for the GCs of interest. In our models at $t=12\,\gyr$ we find that the median masses in \popone\ and \poptwo\ are $0.83\,\msun$ and $0.43\,\msun$, respectively. The ranges in masses overlap marginally due to the binaries in \poptwo. 
GCs are old stellar populations with a small range in ages. Thus, defining the two populations based on the $\Lto$ also ensures that very similar mass ranges are selected in the populations independent of the model. In addition, for real GCs, it is trivial to transfer these definitions to any magnitude system simply remembering the conversion from $L$ to magnitude.  
These benefits will be apparent when we consider real GCs (\autoref{S:observational_results}). 
\begin{figure*} 
\plotone{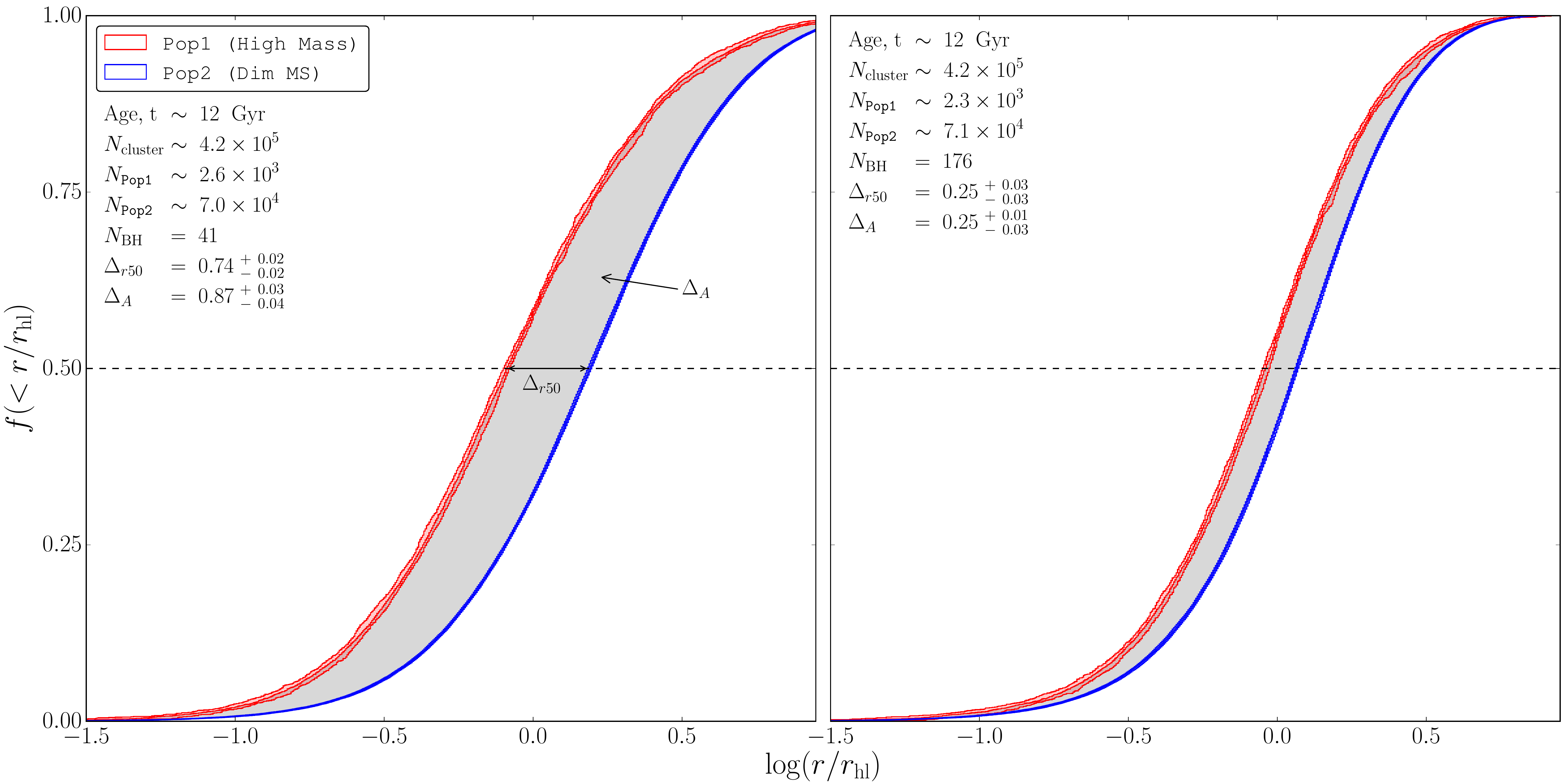}
\caption{Example cumulative distributions for the projected radial locations in the cluster for \popone\ and \poptwo\ objects (single and binary stars; \autoref{S:population_selection}). Left and right panels show examples from clusters containing $41$ and $176$ BHs at $t=12\,\gyr$ (models no. 6 and 22; \autoref{T:modellist}). Red and blue histograms denote \popone\ and \poptwo, respectively. In each case, the red and blue shaded regions denote the range in statistical fluctuations based on $25$ realizations of 2D projections for the same radial distribution of stars in the cluster. Clearly, the higher-mass population \popone\ is more centrally concentrated than the lower-mass population \poptwo. The measures of mass segregation ($\Deltarfifty$ or $\Deltaa$) between \popone\ and \poptwo\ depend on the number of retained BHs. For example, the model cluster shown in the right panel containing a larger $\nbh$ shows a clear quenching of $\Delta$ between \popone\ and \poptwo\ relative to the cluster shown in the left panel containing fewer $\nbh$.  
}
\label{f:2}
\end{figure*}

We want to mention that other populations could be compared as well. For example, recent studies \citep{2016MNRAS.462.2333P,2016ApJ...833..252A} have used BSs and stars on the upper MS as two populations. Both populations are luminous and BSs are expected to be more massive than stars on the MS. However, while these choices are sufficient to demonstrate the anti-correlation between mass segregation and $\nbh$, stochastic noise due to the small size of the BS population can limit the accuracy of any measurement of $\Delta$, especially since it is also expected that high BH retention should decrease BS formation efficiency \citep[e.g.,][]{1993Natur.364..423S}. In contrast, giants and low-mass MS stars are both plentiful in a typical GC. As an example, the model represented in \autoref{f:1} (model no. 6 in \autoref{T:modellist}) at $t=12\,\gyr$ contains
$\sim3\times10^3$, $\sim7\times10^4$, and $\approx36$ \popone, \poptwo, and BSs, respectively.
In addition, BSs by definition changed mass at some unknown past epoch, whereas, the majority of both \popone\ and \poptwo\ defined above are normal, undisturbed stars. Hence, mass segregation between stellar populations defined as above is easier to interpret compared to populations involving BSs in past studies.

\subsection{Quantifying Mass Segregation}\label{S:delta_def}
Having chosen two distinct populations, we now define two similar methods for quantifying $\Delta$ between \popone\ and \poptwo\ using the cumulative distribution functions (CDF) of the $2$D projected cluster-centric positions (\autoref{f:2}). Our fiducial parameter is defined to be the difference between the projected cluster-centric distances at the medians of the two cumulative distributions described above, normalized by the $\rhl$:

\begin{equation}
\label{e:deltar50}
\Deltarfifty=\frac{r_{50,2} - r_{50,1}}{r_{\rm{hl}}}. 
\end{equation}
Here, $r_{50,i}$ denotes the median of the CDF for the projected locations of the $i^{\rm{th}}$ population. 

Simply to ensure that our results are not sensitive to the exact definition of $\Delta$, we adopt another equivalent definition following \citet{2016ApJ...833..252A}, namely, the normalized area between the CDFs for the two populations:
\begin{eqnarray}
\label{e:deltaa}
\Deltaa & = & \int_{r_{\rm{min}}}^{r_{\rm{max}}} \frac{dr}{r_{\rm{hl}}} \biggr[f_{1}\biggr(\frac{r}{r_{\rm{hl}}}\biggr) - f_{2}\biggr(\frac{r}{r_{\rm{hl}}}\biggr)\biggr] \nonumber\\
 & = & \frac{A_1 - A_2}{r_{\rm{hl}}}.
\end{eqnarray}

Here, the integration is performed over the full range in projected cluster-centric distances, $f_i$ denotes the CDF for population $i$, and $A_i$ denotes the area under $f_i$.  

While $\Deltaa$ samples all parts of the CDFs and may be more robust, especially when one of the populations does not contain a large enough number of stars, we find that in our models and our stellar populations, $\Deltarfifty$ and $\Deltaa$ show little difference in precision and level of fluctuations between realizations (for the 2D projections). 
Hence, while we present both $\Deltaa$ and $\Deltarfifty$ for all our models in \autoref{T:results}, we show the simpler-to-compute $\Deltarfifty$ in subsequent figures. Note that the same figures could also be made for $\Deltaa$ and they would show essentially identical correlations.   

\begin{figure*}
\plotone{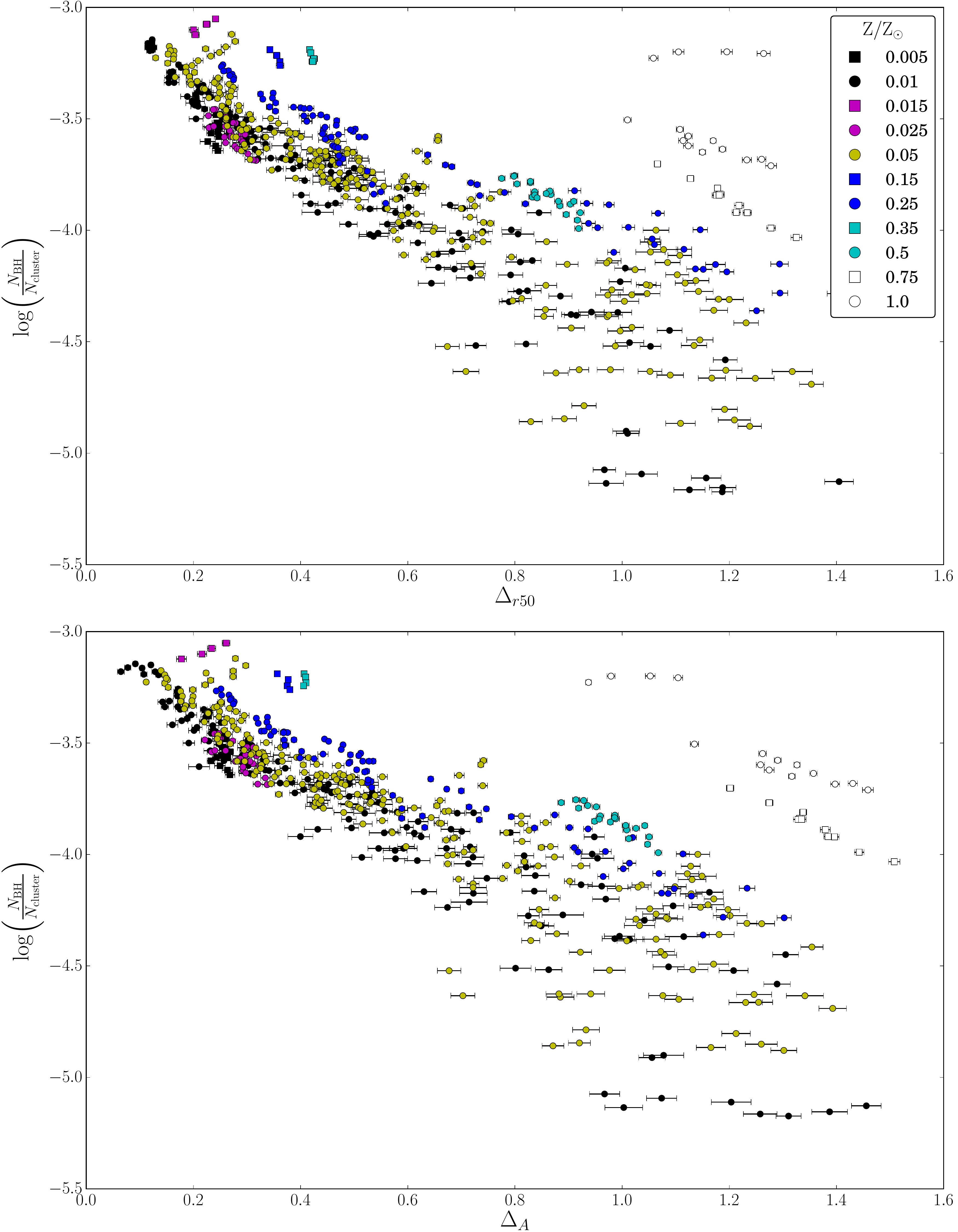}
\caption{Number of retained BHs per cluster star $\nbh/\ncluster$ vs $\Deltarfifty$ (top) and $\Deltaa$ (bottom) for all model snapshots with $t\geq9\,\gyr$ (\autoref{S:delta_def}). Each point represents the mean of $\Deltarfifty$ ($\Deltaa$) calculated using Eq.\ \ref{e:deltar50} (\ref{e:deltaa}) from $25$ realizations of 2D projections of the radial positions of all stars in a model cluster snapshot. The error bars represent the standard deviation within these realizations. Point shape and color denote different stellar metallicities (shown in the label). A clear anti-correlation between $\nbh/\ncluster$ and $\Deltarfifty$ ($\Deltaa$) is apparent, especially if models of particular metallicities are considered. In addition, there is a metallicity effect: the higher the metallicity, the higher the $\Deltarfifty$ (or $\Deltaa$) for any given $\nbh/\ncluster$. As $\metal$ increases, BH masses decrease. Thus, to effect the same level of quenching of $\Delta$, a higher $\nbh/\ncluster$ is needed. 
}
\label{f:3}
\end{figure*}

%%%%%%%%%%%%%%%%%%%%%%%%%%%%%%%%%%%%%%%%%%%%%%%%%%%%%%%%%%%%%%%%%%%%%%%%%%%%%%%%%%%%%%%%%%%%%%%%%
%
\section{Number of Black Holes and Mass Segregation: Models} \label{S:model_results}
%
%%%%%%%%%%%%%%%%%%%%%%%%%%%%%%%%%%%%%%%%%%%%%%%%%%%%%%%%%%%%%%%%%%%%%%%%%%%%%%%%%%%%%%%%%%%%%%%%%

In this section we study the correlation between $\nbh$ in a model cluster and $\Delta$ measured using \popone\ and \poptwo\ stars in that model (\autoref{S:population_selection}; \autoref{f:3}). Note that in terms of causality, retention of BHs quenches $\Delta$. However, since $\Delta$ can be measured in a real GC, and our goal is to infer $\nbh$ from $\Delta$, we treat $\Delta$ as the independent variable and $\nbh$, normalized by the total number of cluster stars, as the dependent variable. \autoref{f:3} contains results from a total of 13,850 independent realizations of sky-projected clusters. Each data point corresponds to a snapshot of an independent model between $9$ and $12\,\gyr$. The markers and error bars denote the median and standard deviation based on $25$ independent 2D projections performed on each snapshot.

Several interesting trends emerge. Both measures of mass segregation, $\Deltarfifty$ and $\Deltaa$, show the same strong anti-correlation with $\nbh/\ncluster$, i.e., the higher the $\nbh/\ncluster$, the lower the $\Delta$. On top of this general trend, there is also an apparent trend dependent on the stellar metallicity. The higher the $\metal$, the lower the mass of the BHs produced. As a result, to quench $\Delta$ to the same level, a higher-$\metal$ cluster needs higher $\nbh/\ncluster$ compared to a lower-$\metal$ cluster. Other parameters such as the initial binary fraction $f_b$, natal kick distribution, and cluster age contribute to the spread in the observed trend and can be marginalized in the overall calibration involving only $\Delta$ and $\nbh/\ncluster$. For example, a higher natal kick distribution simply ejects more BHs from the cluster at birth, leading to a lower $\nbh/\ncluster$ at a later time $t\geq9\,\gyr$. Similarly, variations in $f_b$ slightly modify the relative masses of the visible populations and those of the retained BHs \citep[typically single due to frequent dynamical encounters, wind mass loss, and supernova; e.g.,][]{2017ApJ...834...68C} and contribute to the spread in the anti-correlation between $\nbh/\ncluster$ and $\Delta$.

Note that statistical fluctuations due to independent projections of the snapshots are of similar magnitude for both $\Deltarfifty$ and $\Deltaa$. This ensures that fluctuations in the measure of mass-segregation are realistic and depend simply on the viewing angle to a particular cluster. These fluctuations are not due to lack of robustness in our defined measure of mass segregation. This is only expected, since both our defined stellar populations contain a sufficiently large number of stars, and as a result, their CDFs do not suffer from noise arising from small numbers. 

These models now give us the power to create a statistical calibration relating $\Delta$ ($\Deltarfifty$ or $\Deltaa$) to $\nbh/\ncluster$ in real clusters. Note that in our definitions, both quantities of interest, $\nbh/\ncluster$ and $\Delta$, are dimensionless. Moreover, the former and latter relate to a real cluster through the total mass and the half-light radius, both of which are easily observable within some uncertainties related to the mass-to-light ratio. To properly consider the large realistic spreads in the parameter space depending on the cluster initial properties, we use a non-parametric gaussian kernel density estimation (KDE) to represent the probability distribution function (PDF) in the $\Deltarfifty$--$\nbh/\ncluster$ (alternatively, $\Deltaa$--$\nbh/\ncluster$) plane instead of a deterministic best-fit approach. In later sections where we use this KDE to predict $\nbh/\ncluster$ in real clusters, we further restrict ourselves to models with $\metal$ close to that of the respective observed clusters. As we will see, in real clusters we also need to limit model results to the maximum cluster-centric distance available in the ACS survey for individual clusters.

%%%%%%%%%%%%%%%%%%%%%%%%%%%%%%%%%%%%%%%%%%%%%%%%%%%%%%%%%%%%%%%%%%%%%%%%%%%%%%%%%%%%%%%%%%%%%%%%%
%
%\section{Data Reduction and Measuring Mass Segregation for Observed Clusters}
\section{Measuring Mass Segregation for Observed Clusters}
\label{S:observational_results}
%
%%%%%%%%%%%%%%%%%%%%%%%%%%%%%%%%%%%%%%%%%%%%%%%%%%%%%%%%%%%%%%%%%%%%%%%%%%%%%%%%%%%%%%%%%%%%%%%%%
We now proceed to infer $\nbh$ from measured $\Delta$ in real GCs using calibrations obtained from models. We use the ACS Survey for MWGCs \citep{2007AJ....133.1658S} for the stellar data. Compiled using the wide-field channel of the \textit{Hubble Space Telescope's} Advanced Camera for Surveys (ACS), this resource consists of an exhaustive catalog of stars within the central $4\arcmin\times4\arcmin$ of 65 MWGCs. The catalog exists as an online database of stellar coordinates and calibrated photometry in the ACS VEGA-mag system \citep{2005PASP..117.1049S}. The database may be accessed publicly at \url{http://www.astro.ufl.edu/~ata/public\_hstgc} and its construction is fully detailed in \citep{2008AJ....135.2055A}. We currently limit our analysis to 47\ Tuc, M\ 10, and M\ 22 -- three of four known MWGCs to contain candidate stellar-mass BHs \citep[e.g.,][]{2012Natur.490...71S,2014cxo..prop.4353S,2015MNRAS.453.3918M,2017MNRAS.467.2199B}. Relevant structural properties of these three GCs are listed in \autoref{T:props} for easy reference. While M\ 62 also contains a stellar-mass BH candidate \citep[e.g.,][]{2013ApJ...777...69C}, this cluster is not included in the publicly available ACS survey catalog. 

\begin{deluxetable*}{ccccccccc}
\tabletypesize{\scriptsize}
\tablecolumns{9}
\tablewidth{0pt}
\tablecaption{Selected Properties of the GCs}
\tablehead{
    \multicolumn{2}{c}{Name} & $M_{\rm{cluster}}$ & $\rcobs$ & $\rhl$ & \multicolumn{2}{c}{Harris Metallicity} & \multicolumn{2}{c}{ACS Metallicity} \\
    NGC & Alt & [$10^5 M_{\odot}$]$\tablenotemark{a}$ & [arcmin]$\tablenotemark{b}$ & [arcmin]$\tablenotemark{b}$ & [Fe/H]$\tablenotemark{b}$ & $\metal/\metal_{\odot}$ & [Fe/H]$\tablenotemark{c}$ & $\metal/\metal_{\odot}$}
\startdata
0104 & 47 Tuc & 7.00 & 0.36 & 3.17 & -0.72 & 0.191  & -0.78 & 0.166  \vspace{-0.1cm} \\
6254 & M 10   & 1.00 & 0.77 & 1.95 & -1.56 & 0.0275 & -1.25 & 0.0562 \vspace{-0.1cm} \\
6656 & M 22   & 2.90 & 1.33 & 3.36 & -1.70 & 0.0200 & -1.49 & 0.0324 \\ 
\enddata
\tablenotetext{a}{\citet{2010MNRAS.406.2000M}}
\tablenotetext{b}{Harris catalog \citep[2010 edition]{1996AJ....112.1487H}}
\tablenotetext{c}{\citet{2009ApJ...694.1498M}}
\label{T:props}
\end{deluxetable*}

In contrast to models where all relevant properties of all stars and binaries are known, analyzing real GC data is challenging in several ways. For our purposes, the most important considerations are the following. The measure of mass segregation depends on the radial limit to which stellar data is available. Because of the fixed angular size of the ACS survey, the radial extent for the available data is dependent on the heliocentric distance to the cluster. The CMD can have large scatter and the available database does not include detailed cluster-membership information. Since cluster centers are crowded fields, the completeness of stars depends on the location as well as the magnitude which can introduce biases in our measures of mass segregation. Below we describe how we analyze the ACS survey data to mitigate these challenges.  

\begin{figure*}
\plotone{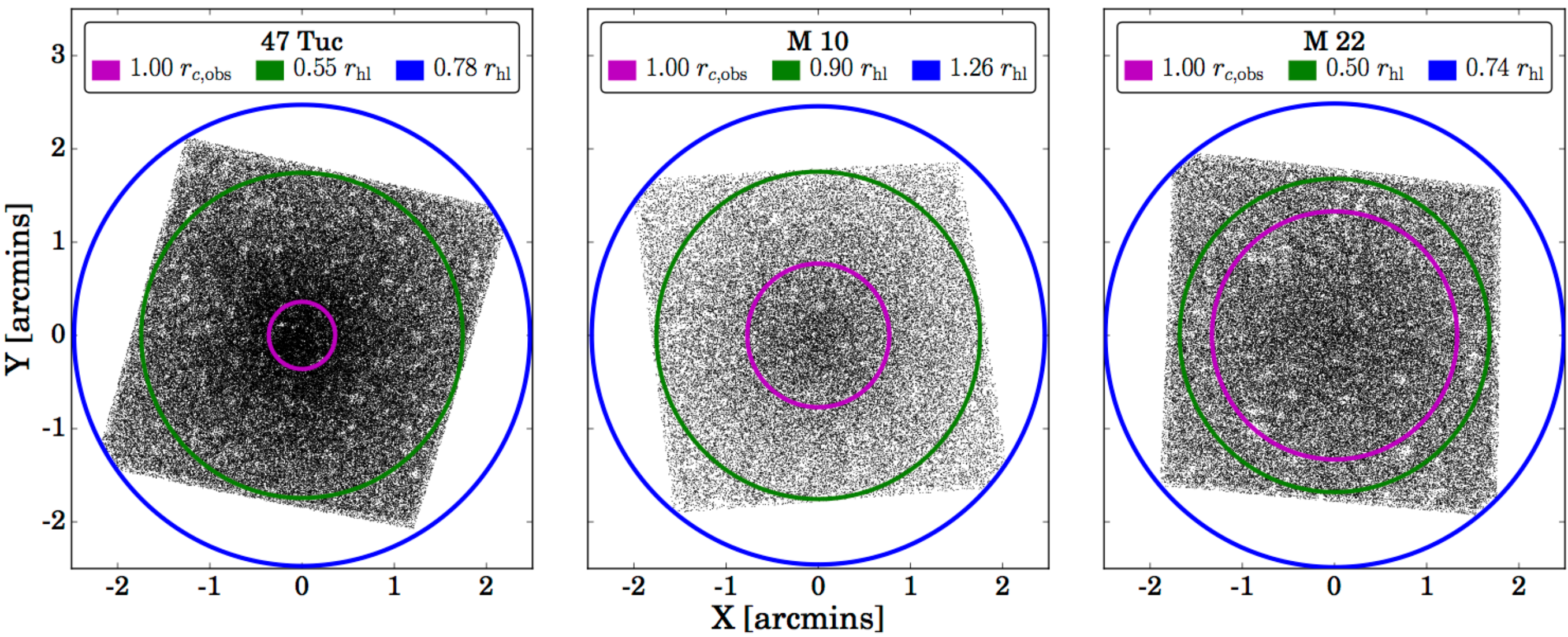}
\caption{Data stamps showing the fields of view for 47\ Tuc (left), M\ 10 (middle), and M\ 22 (right) in the publicly available data from the ACS survey for the MWGCs \citep{2007AJ....133.1658S}. The $X$ and $Y$ coordinates are obtained directly from the ACS catalog after converting the original pixel units into arcminutes \citep{2008AJ....135.2055A}. For reference, observed core radii are shown (magenta) for each GC. Green circles denote the largest radial extent that can be inscribed into the data stamp. For each GC, only stars contained within the green circle are considered for the calculation of mass segregation. The blue circles show the radial extent that can be superscribed to the data stamp. The observed core and half-light radii are adopted from the Harris catalog for MWGCs \citep[2010 edition]{1996AJ....112.1487H}.
}
\label{f:4}
\end{figure*}

\subsection{ACS Field of View and Limiting Radius} \label{S:radlim}
Because the field of view (FOV) of the ACS only covers an approximate $4\arcmin \times 4\arcmin$ rhombus centered on the cluster core (\autoref{f:4}), it is not possible to analyze ACS data all the way out to the tidal radius. This radial limitation on the ACS data will {\em reduce} the measured values of $\Deltarfifty$ and $\Deltaa$ relative to those computed for the radially complete CMC models in \autoref{S:model_results}. 

The available data for each cluster of interest in this paper is shown in \autoref{f:4}. In each case the dots denote the actual stellar positions. In addition, we show the size of the observed core radius ($\rcobs$) and the size of the largest circle ($\rlim$) centered at the cluster center that can be inscribed in the FOV. This allows us to only include stars up to $\rlim/\rhl=0.55$, $0.9$, and $0.5$ for 47\ Tuc, M\ 10, and M\ 22, respectively. In each case, we recalculate $\Deltarfifty$ (and $\Deltaa$) for our models using the same definitions described in \autoref{S:models} but now including only stars up to a projected cluster-centric distance of $r/\rhl_{,\rm{model}} = \rlim/\rhl$. Of course, since each observed cluster has a slightly different value for $\rlim/\rhl$, the measure of mass segregation needs to be recalculated from models separately for each individual cluster imposing the unique radial limit. KDEs created from these custom-calculated $\Deltarfifty$ and $\Deltaa$ along with $\nbh/\ncluster$ are used to estimate the PDF for the expected number of BHs retained in each observed cluster.

While imposing custom radial limits matching the observed clusters reduces the numbers of stars in both \popone\ and \poptwo\ in our models, the resulting increase in statistical scatter is negligible relative to the real spread in $\nbh/\ncluster$ for any given value of $\Deltarfifty$ or $\Deltaa$. This is simply because for all radial limits imposed by the ACS FOV, there is always a large enough number of stars in \popone\ and \poptwo\ as defined in \autoref{S:population_selection}. Still, we test the robustness of our final $\nbh/\ncluster$ estimates by employing several smaller than necessary radial limits for M\ 10, the GC where $\rlim/\rhl$ is the largest.

\subsection{Population Selection} \label{S:popselec2}
\begin{figure*}
\plotone{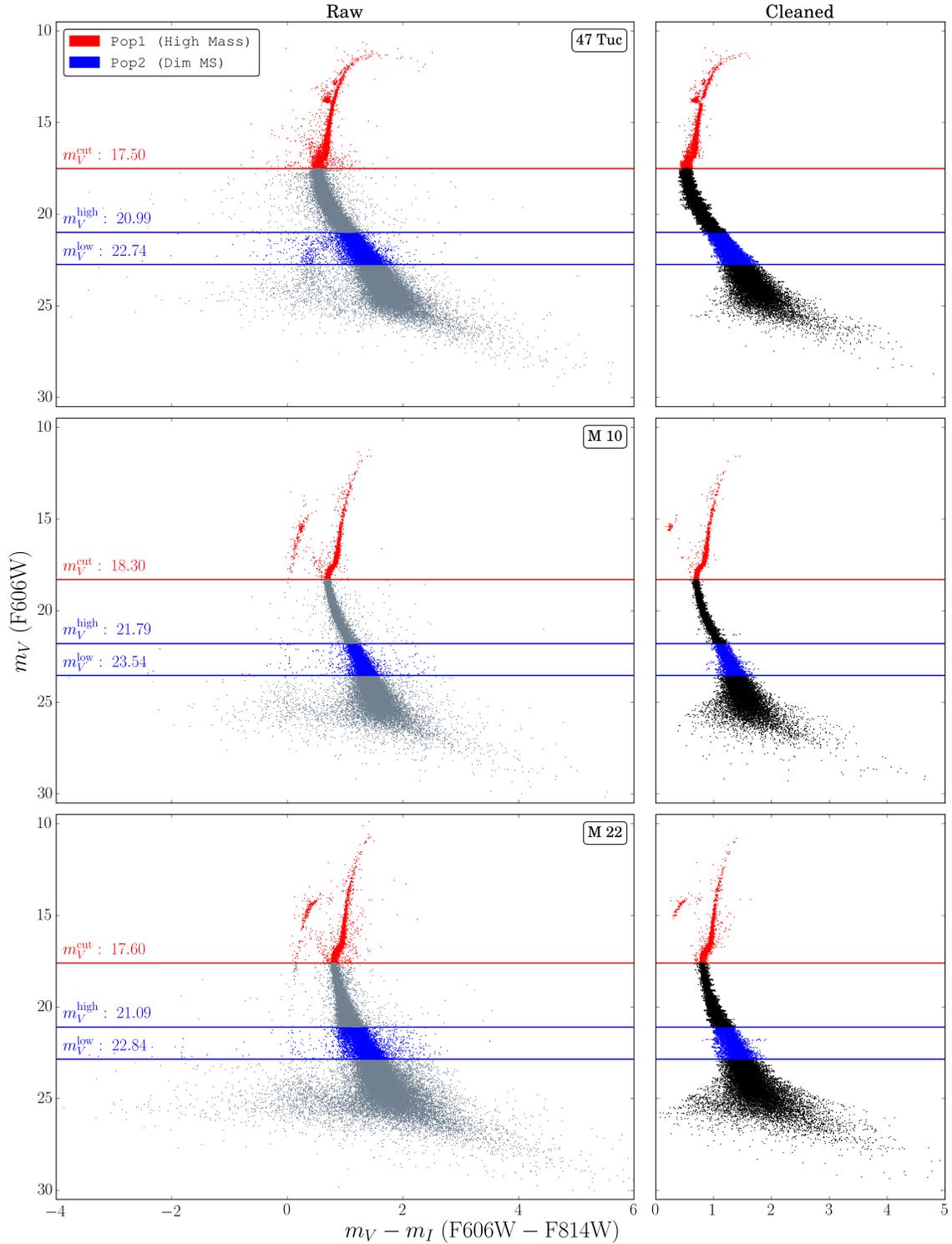}
\caption{Raw  (left) and cleaned (right) CMDs for 47\ Tuc (top), M\ 10 (middle), and M\ 22 (bottom). The raw CMDs are cleaned by discarding all stars outside $2\sigma$ in the color distribution for each bin in the $m_V$ (see \autoref{S:popselec2}). Each dot represents an object (single stars or unresolved binaries from the catalog). Red and blue dots denote \popone\ and \poptwo\ stars in each GC following the same definition used in our models (\autoref{f:1}; Sections\ \ref{S:population_selection}, \ref{S:popselec2}).
}
\label{f:5}
\end{figure*}

Once the radial limits are imposed, population selection proceeds as described in \autoref{S:population_selection} but uses CMDs based on the ACS VEGA-mag photometric system. The two filters used in the ACS GC survey to construct CMDs are F606W (corresponding to the V-band) and F814W (corresponding to the I-band). Since our populations are defined by bounds in the CMD relative to the point of highest temperature on the MS, the definitions are applicable to the GC data with any filter after a simple conversion from luminosity ratios (used in models) to magnitude ratios. We simply need to ascertain the equivalent of $\Lto$ ($m_V^{\rm{cut}}$) for a particular cluster using those filters. 
In order to be consistent in our treatment of the ACS data, $m_V^{\rm{cut}}$ is determined in the following way. We select the relevant portion of the CMD along the V-band axis that contains the turnoff -- for the clusters we consider, this range is set between $m_V = 15$ and $20$. We split this range into 50 bins of 0.1-mag width. $m_V^{\rm{cut}}$ is then defined to be the midpoint of the bin with the lowest mean color, $\langle m_V-m_I \rangle$. All stars above this point are included in \popone. The 1/25th and 1/125th luminosity bounds in the definition for \poptwo\ correspond to bounds in V-band magnitudes $m_V^{\rm{cut}} + 3.49$ and $m_V^{\rm{cut}} + 5.24$, respectively. The resulting population bounds are shown in \autoref{f:5}.

To be more conservative in our populations for real clusters we employ a further cut.  
The raw CMDs made from the ACS survey's online catalog have large spreads along the MS as well as the giant branch, with several outliers (\autoref{f:5}). A sizable number of outliers have high photometric errors. Some of these stars may also simply not be cluster members. Note that for our purposes it is alright to undersample stars in both populations so long as no radial bias is introduced. Keeping this in mind we outline below how we proceed to cleanup the raw data.

The traditional process to clean up the CMD of a cluster is to employ a cut on the photometric error and measurement quality. However, as warned in \citet{2008AJ....135.2055A}, cuts based strictly on the quality estimates may introduce radial biases. For example, due to crowding in regions of high stellar number density like the core, point-spread functions (PSFs) of nearby stars -- especially dim neighbors to a bright giant -- are not always separately resolvable. This radially biases the relative locations for giants and low-mass MS stars, essentially artificially increasing the measured $\Delta$. Crowding introduces another bias in photometric error. Overlapping PSFs of highly-proximate stars cannot be individually fitted as well as those in isolation. Hence, stars in the core, especially dim ones, suffer from preferentially high measurement uncertainties. Any quality cut would therefore preferentially select the brighter and more isolated stars, again introducing bias leading to an artificial increase in the measured $\Delta$. Since the primary measurable of interest for this study is the relative radial distributions of two populations highly different in their typical magnitudes, we cannot use the above-mentioned traditional methods.

\begin{figure*}
\plotone{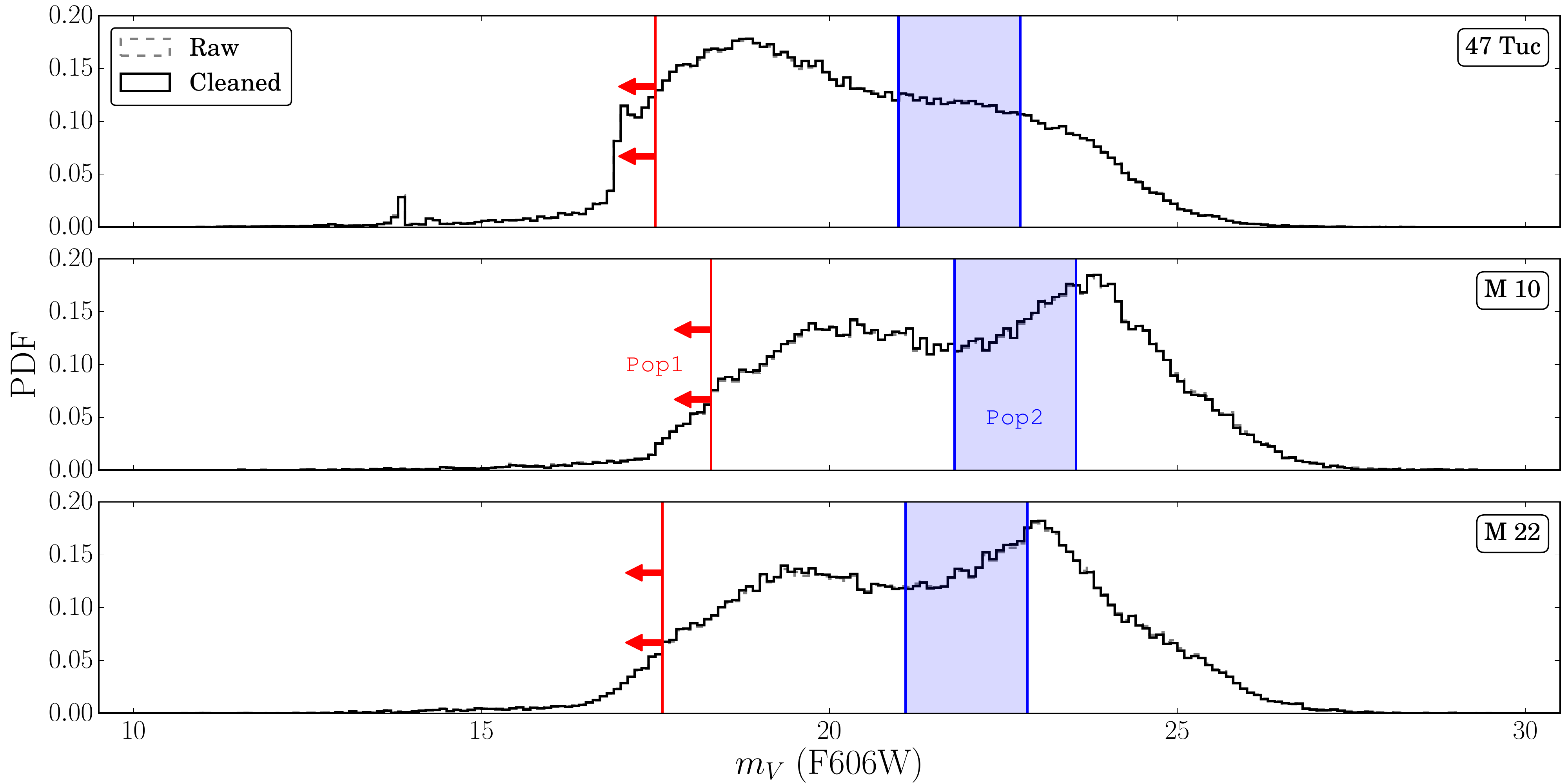}
\caption{Luminosity functions (LF) in the apparent VEGA-mag photometric system corresponding to the raw (dashed) and cleaned (solid) CMDs (\autoref{f:5}; \autoref{S:popselec2}) for 47\ Tuc (top), M\ 10 (middle), and M\ 22 (bottom). Left of the red vertical lines, and the blue shaded region denote the LFs for \popone\ and \poptwo\ stars. The nearly identical LFs between the raw and cleaned populations, especially in ranges corresponding to \popone\ and \poptwo\ indicate that our cleanup step did not introduce any significant bias in population selection. 
}
\label{f:6}
\end{figure*}

\begin{figure*}
\plotone{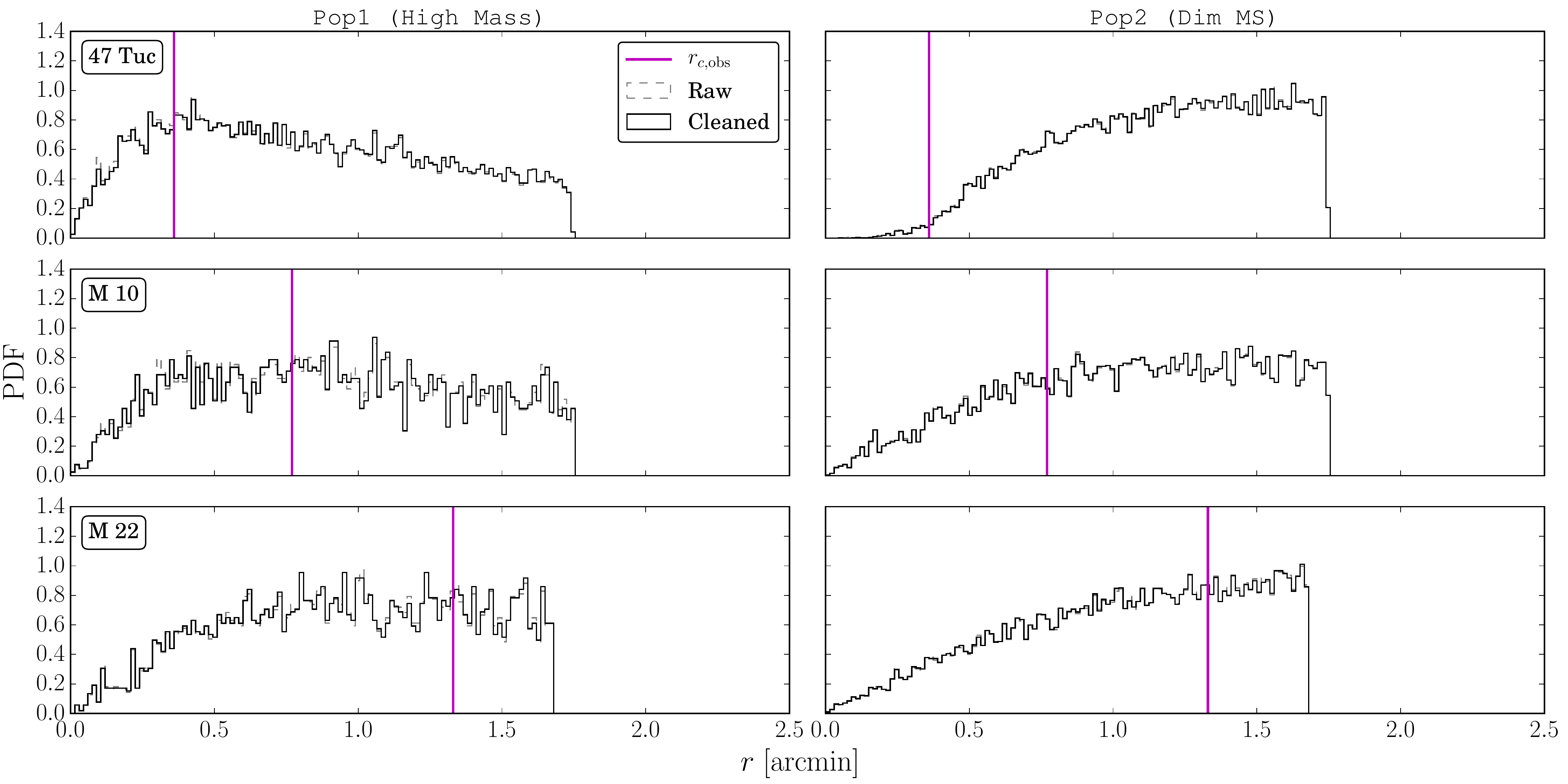}
\caption{Radial distribution for \popone\ (left) and \poptwo\ (right) stars corresponding to the raw (dashed) and cleaned (solid) CMDs (\autoref{f:5}; \autoref{S:popselec2}). The vertical magenta lines indicate the core radius for reference. The distributions' sharp ends are due to the radial limit imposed by the ACS field of view. The nearly identical radial distributions for \popone\ and \poptwo\ stars between the raw and cleaned populations indicate that our cleanup step did not introduce any significant radial bias in the two populations.    
}
\label{f:65}
\end{figure*}

\begin{figure*}
\plotone{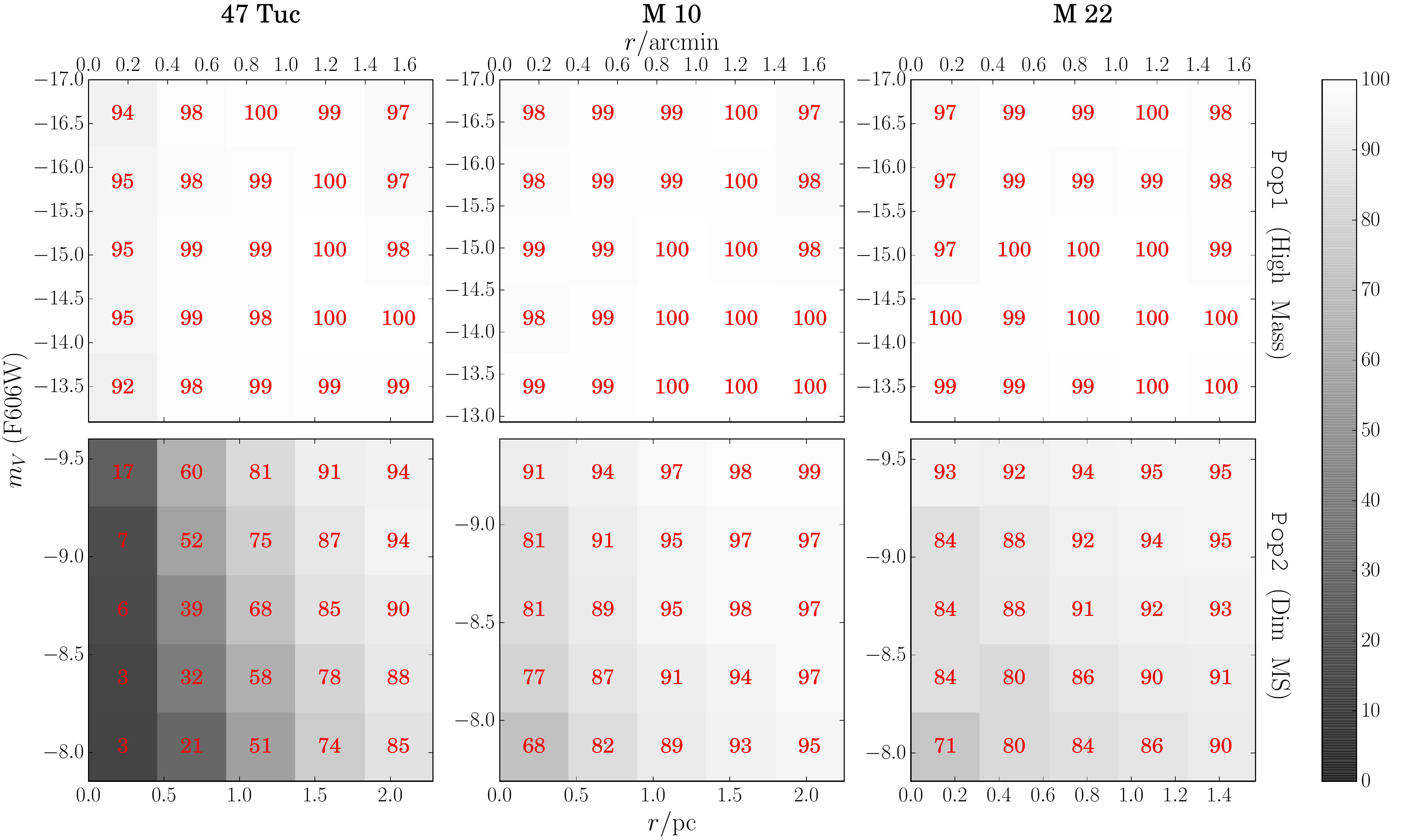}
\caption{Completeness map as a function of the apparent instrumental magnitude and cluster-centric radial position for \popone\ (top) and \poptwo\ (bottom) for 47\ Tuc (left), M\ 10 (middle), and M\ 22 (right). The completeness fraction is shown in grey-scale. For easy reference, the average completeness fractions are also shown in red in $5\times5$ bins in each panel. As expected, the completeness fraction for giants is nearly 100\% in all bins. However, the dimmer low-mass MS stars have much lower completeness fractions, in particular, close to the center of the GCs. The relative completeness fractions are taken into account to debias the calculation of $\Deltarfifty$ and $\Deltaa$ (\autoref{S:completeness}; \autoref{f:8}).  
}
\label{f:7}
\end{figure*}

For our purposes we employ an empirical scheme based on the proximity of the stars from the nominal location of the CMD in the magnitude-color space.
We first sort all stars (left panels, \autoref{f:5}) into bins of $\delta m_V=0.1$. We create a PDF for the stars in each $m_V$ bin along the color axis. We find the locations of the mode and mode-centric $2\sigma$ limits in color for each $m_V$ bin. For each $m_V$ bin we exclude all stars outside the $2\sigma$ color limits to create a cleaner version of the CMD (right panels, \autoref{f:5}). We calculate $\Delta$ ($\Deltarfifty$ and $\Deltaa$) using the stars in the cleaned CMD and using the same definitions of \popone\ and \poptwo\ stars as described above.   

Note that our main goal is to ascertain that our CMD-cleaning stage does not preferentially exclude or include stars of either \popone\ or \poptwo. For all of the three GCs we consider, the luminosity functions (LFs) are almost identical before and after the cleaning step, especially in the regions of our \popone\ and \poptwo\ for these GCs (\autoref{f:6}). This gives us confidence that the CMD cleaning stage cuts down both populations proportionally without introducing any bias in location in the cluster and $m_V$ or $m_V-m_I$. 
We also ascertain that the radial distributions of \popone\ and \poptwo\ stars before and after CMD cleaning remain unchanged (\autoref{f:65}), further establishing that the CMD cleaning stage does not introduce any biases. 

\subsection{Correcting for Completeness} \label{S:completeness}

Due to crowding, photometric completeness decreases with increasing magnitude and decreasing radial distance from the center. In other words, a dimmer stellar population will be less complete than brighter stars near the center. Difference in completeness can artificially increase the measured 
$\Delta$ if not corrected properly. To correct for these biases, we first determine the completeness as a function of magnitude and radial distance using the ACS catalog's artificial star files \citep[][see their Section\ 6 for a full description]{2008AJ....135.2055A}. In short, \citet{2008AJ....135.2055A} generate $10^5$ PSFs of artificial stars drawn from the LF between $m_v=-17$ and $-5$ (instrumental magnitudes) and inject them on top of the raw fields for each cluster, spaced far enough apart to avoid overlap and distributed according to a spatial density that is flat within the core and scales as $r^{-1}$ outside. They then try to recover the injected artificial stars using the same star-finding procedure from the raw images. If the recovered positions and V,I-band magnitudes are within 0.5 pixel and 0.75 mag of the injected values, they count the stars as successfully `recovered'.

We use the radial location $r$ and magnitude $m_V$ of all recovered stars to construct a bivariate PDF $\nrec(r,m_V)$ using a gaussian KDE. Similarly, we create another PDF for all injected stars $\ninject(r,m_V)$. The completeness fraction is then simply given by 
\begin{equation} \label{e:8}
C(r,m_V) = \frac{\nrec(r,m_V)}{\ninject(r,m_V)}.  
\end{equation}
\autoref{f:7} shows the distribution of $C(r,m_V)$ as a function of $r$ and $m_V$ for the relevant range in $r$ and the two relevant ranges in $m_V$ for \popone\ and \poptwo\ stars for all three GCs of interest. 

Note that the completeness decreases significantly for \poptwo\ as $r$ decreases for all clusters. Completeness for \popone\ stars also shows a similar trend but being bright, they are not affected by crowding as much as the \popone\ stars. To correct for this difference in completeness between the two populations we under-sample the population with higher completeness based on the relative completeness fractions as follows.   

We divide the relevant parameter space in $r$ and $m_V$ into $100\times100$, a total of $10^4$ bins. We calculate the completeness $C_{i,j}$ at the midpoint of each bin. We then find the minimum in $C_{i,j}$ and define a probability of sampling as 
\begin{equation}
W_{i,j} = \frac{\min (C_{i,j})}{C_{i,j}}.
\end{equation}
For example, for M\ 10, $\min(C_{i,j})\approx68\%$ for the \poptwo\ stars, corresponding to the most centrally located and least bright bin.
Stars in a bin with $C_{i,j}=99\%$ would then have a probability of selection $W_{i,j} = 68/99$.

\begin{figure*}
\plotone{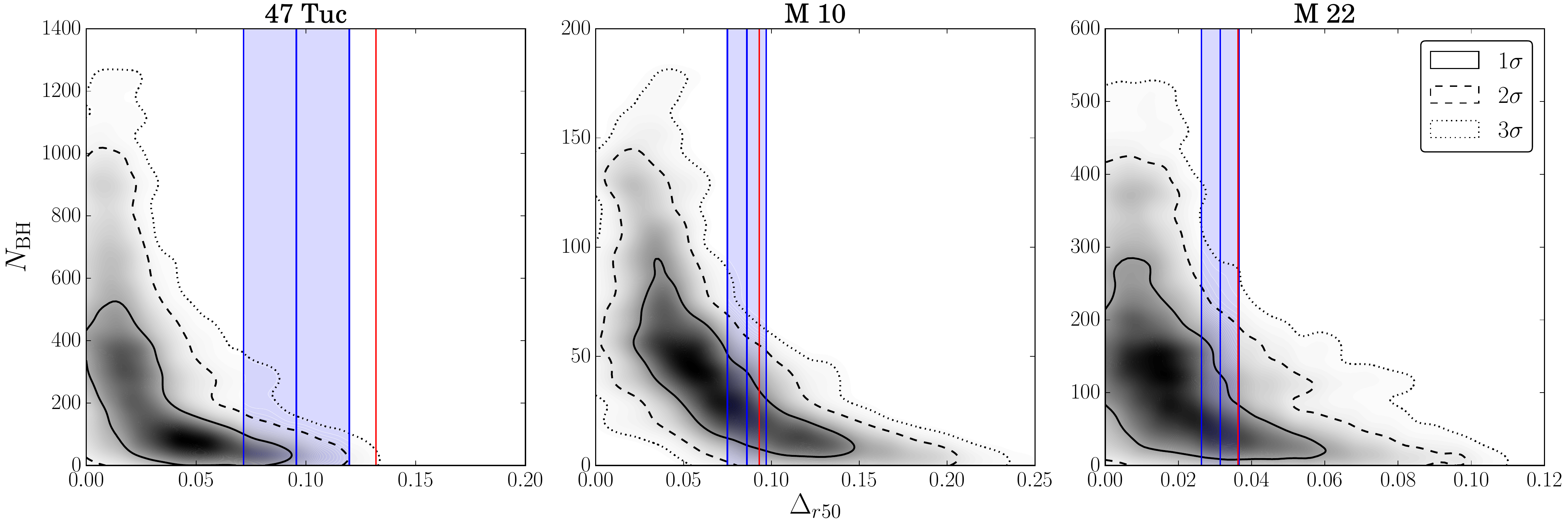}
\caption{Bivariate PDFs for $\nbh$ vs $\Deltarfifty$ for 47\ Tuc, M\ 10, and M\ 22 from models adjusted using the same limiting radius $\rlim/\rhl$ as available for the real GCs (see \autoref{S:radlim}). Black solid, dashed, and dotted contours denote $1$, $2$, and $3\sigma$ for the model PDFs. The vertical blue lines and shaded regions denote the mean and $1\sigma$ for the $\Deltarfifty$ after all corrections (\autoref{S:observational_results}). The vertical red lines denote $\Deltarfifty$ before completeness correction for reference (\autoref{S:completeness}). Note that, to remain consistent with the observed metallicities of these GCs only models with $0.01\leq\metal/\metal_\odot\leq 0.25 Z$ have been used to create the bivariate PDFs from models.  
}
\label{f:8}
\end{figure*}

Using these sampling probabilities, stars are randomly selected from both populations and $\Deltarfifty$ and $\Deltaa$ are calculated using those selected stars. To understand the levels of statistical fluctuations we repeat this exercise $1000$ times to obtain distributions for these quantities. This same process is repeated for all three chosen GCs. The mean and standard deviations for these quantities are given in \autoref{T:results} for each of these clusters. \autoref{f:8} shows $\Deltarfifty$ before and after correcting for completeness -- in each case, completeness correction has an impact. The results are very similar for $\Deltaa$.

\begin{figure*}[htb!]
\plotone{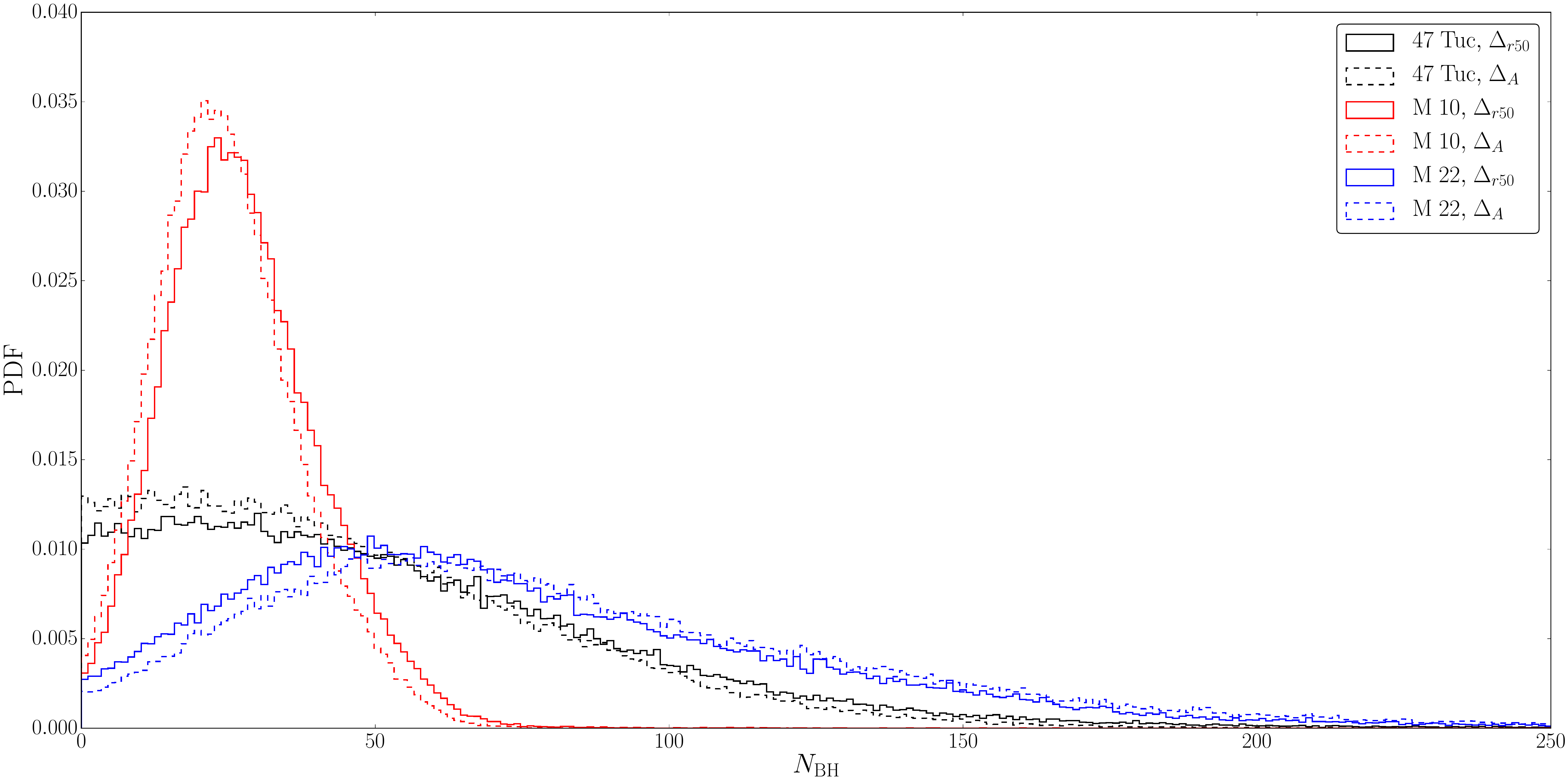}
\caption{Distributions of the retained number of BHs in each GC predicted from the observed levels of mass segregation. Solid and dashed markings denote distributions obtained using $\Deltarfifty$ and $\Deltaa$, respectively. Black, red, and blue denote the $\nbh$ distributions retained in 47\ Tuc, M\ 10, and M\ 22.  
}
\label{f:9}
\end{figure*}

%%%%%%%%%%%%%%%%%%%%%%%%%%%%%%%%%%%%%%%%%%%%%%%%%%%%%%%%%%%%%%%%%%%%%%%%%%%%%%%%%%%%%%%%%%%%%%%%%
%
\section{Predicting the Number of Retained Black Holes in Observed GCs} \label{S:nbh_predictions}
%
%%%%%%%%%%%%%%%%%%%%%%%%%%%%%%%%%%%%%%%%%%%%%%%%%%%%%%%%%%%%%%%%%%%%%%%%%%%%%%%%%%%%%%%%%%%%%%%%%

We now proceed to derive PDFs for $\nbh$ retained in 47\ Tuc, M\ 10, and M\ 22, as inferred from the measured $\Deltarfifty$ and $\Deltaa$. Our numerical models are used to individually calculate $\Deltarfifty$ and $\Deltaa$ using the proper radial limit (\autoref{S:radlim}) for each cluster. The radial limits used for each individual GC are given in \autoref{T:results} for reference. Furthermore, we restrict ourselves to models with $0.01\leq\metal/\metal_\odot\leq0.25$ in order to be roughly consistent with the low metallicities of the three GCs of interest. We construct individual bivariate distributions for $\Delta$ and $\nbh/\ncluster$ from these low-$\metal$ models using gaussian KDEs. These distributions are then used to infer the expected number of retained BHs in each GC in the following way. 

For each GC we randomly sample $10^5$ values of $\Deltarfifty$ (or $\Deltaa$) from a normal distribution with the mean and standard deviation calculated in \autoref{S:completeness}. For each of these draws of $\Deltarfifty$ (or $\Deltaa$) we randomly sample one value of $\nbh/\ncluster$ from the bivariate KDE obtained from the models using the appropriate radial limits corresponding to the GC in question requiring that the draw from the KDE has $\Deltarfifty$ (or $\Deltaa$) within $1\%$ of the sampled value for the GC. Thus for each observed GC we obtain $10^5$ values of $\nbh/\ncluster$ based on the measured $\Deltarfifty$ (or $\Deltaa$). We estimate $\nbh$ by assuming $M/L=2$ and that $\ncluster=M_c/0.5$, i.e., the average stellar mass is $0.5\,\msun$.\footnote{Of course this assumption can easily be adjusted. The actual quantity calculated is $\nbh/\ncluster$. Hence, the keen reader can easily use a better estimate of $\ncluster$ in the future and recalculate $\nbh$ for each GC. This simply will shift these distributions by a multiplying factor.   
}

\autoref{f:9} shows the distributions for the predicted retained $\nbh$ in each GC we analyzed, the actual numbers are also listed in \autoref{T:results}. The $\nbh$ distributions for the GCs all have rather large spreads. In all cases the peaks of the distributions are within $\nbh=50$. However, for 47\ Tuc and M\ 22, the maximum $\nbh$ can be significantly larger, namely, $\sim200$ and $\sim300$, respectively. The maximum number of retained BHs in M\ 10 is a little below $\sim100$. In all three GCs we cannot strictly exclude $\nbh=0$, although, zero retention is outside $1\sigma$ (barely for 47\ Tuc) of the distributions for all three GCs. In fact, it is actually hard to rule out zero retained BHs using a method that tracks dynamical effects such as mass segregation in a cluster since below a certain $\nbh/\ncluster$ the presence of BHs likely becomes dynamically unimportant. Thus, a sufficiently small number of retained BHs may not be easily distinguishable from zero BHs in a cluster using mass segregation as a probe. Of course, if the candidate BHs discovered in these GCs truly are BHs, then zero retained BHs is excluded by direct observation. Both measures of mass segregation, $\Deltarfifty$ and $\Deltaa$, give very similar results, while the distributions obtained using $\Deltaa$ are marginally narrower compared to those obtained using $\Deltarfifty$ (\autoref{f:9}).   

The flat distribution for $\nbh$ obtained 
for 47 Tuc (\autoref{f:9}) requires special mention since this illustrates the type of observation that can 
best improve the estimate of $\nbh$ retained in a real cluster using $\Delta$. The flat distribution arises primarily from the low completeness levels for fainter and more centrally concentrated stars in 47 Tuc. Our statistical framework naturally takes this into account and the required undersampling (\autoref{S:completeness}) essentially increases statistical fluctuations in the estimated measure of $\Delta$ (see, e.g., the spread in measured $\Deltarfifty$ for 47 Tuc in \autoref{f:8}). This leads to a flat PDF for the estimated $\nbh$. The only way to be able to predict $\nbh$ better using $\Delta$ as a predictor is to improve the completeness. 

One unavoidable challenge in using this method to predict the retained $\nbh$ in a GC is of course the availability and quality of data. In particular, the radial limit to which the data exists can in principle be crucial, since the larger the radial limit, the higher should be the measured $\Delta$ (\autoref{S:radlim}). While we have mitigated this challenge by limiting our reference model data also to the same limiting radius, we wanted to test how the quality of these predictions change depending on the adopted radial limits. 
Of the three clusters in our sample, M\ 10 has the largest $\rlim/\rhl$. 
Thus, M\ 10 is an ideal GC for a controlled robustness study. While data exists up to $\rlim/\rhl=0.9$ for M\ 10, we repeat the whole process with several smaller $\rlim/\rhl$ values down to $\rlim/\rhl=0.5$. Reducing the radial extent increases completeness corrections as well as decreases the numbers of member stars in both populations. However, since we have defined the stellar populations in such a way that the member number in each group is typically large, we find that artificially limiting $\rlim/\rhl$ does not significantly change our predicted BH numbers for M\ 10. The peaks of the distributions for $\nbh$ are close to each other for all tried $\rlim/\rhl$, though the peak tends to decrease slightly with radial limit, particularly for $\rlim/\rhl\leq0.6$. We also see a slow increase in the $1\sigma$ spreads and $2\sigma$ max with decreasing $\rlim/\rhl$ (\autoref{T:results}). This, together with the similarity in the predicted $\nbh$ using $\Deltaa$ or $\Deltarfifty$, bolsters our belief that the predicted retained $\nbh$ distributions are quite robust.

\floattable
\begin{deluxetable*}{cccccccccc}
\tabletypesize{\scriptsize}
\tablecolumns{10}
\tablewidth{0pt}
\tablecaption{Predicted Numbers of Retained BHs}
\tablehead{
   \multirow{2}{*}{Cluster} & Radial Limit & \multicolumn{2}{c}{Population Size} &
   $\Deltarfifty$ & \multicolumn{2}{c}{$\nbh$} & $\Deltaa$ & \multicolumn{2}{c}{$\nbh$} \vspace{-0.07cm} \\
   & $\rlim/\rhl$ & $N_{\popone}$ & $N_{\poptwo}$ & $\pmsigma$ & $\pmsigma$ & Max ($+2\sigma$) & $\pmsigma$ & $\pmsigma$ & Max ($+2\sigma$)}
\startdata
47 Tuc                & 0.55 &  147 &  264 & $0.0957 \pm 0.0241$ & $19_{-17}^{+51}$ & 153 & $0.0790 \pm 0.0147$ & $16_{-16}^{+46}$ & 124 \\
M 10                  & 0.9  &  828 & 2705 & $0.0860 \pm 0.0111$ & $24_{-12}^{+14}$ &  54 & $0.0656 \pm 0.0074$ & $21_{-10}^{+13}$ &  49 \\
M 22                  & 0.5  & 1317 & 4895 & $0.0314 \pm 0.0051$ & $49_{-34}^{+50}$ & 174 & $0.0215 \pm 0.0032$ & $54_{-35}^{+56}$ & 195 \\ \hline
\multirow{4}{*}{M 10} & 0.8  &  744 & 2330 & $0.0705 \pm 0.0103$ & $23_{-12}^{+14}$ &  54 & $0.0576 \pm 0.0068$ & $20_{-12}^{+10}$ &  46 \\
                      & 0.7  &  660 & 1933 & $0.0548 \pm 0.0102$ & $22_{-12}^{+16}$ &  57 & $0.0441 \pm 0.0065$ & $17_{- 9}^{+14}$ &  49 \\
                      & 0.6  &  564 & 1552 & $0.0407 \pm 0.0097$ & $20_{-13}^{+19}$ &  65 & $0.0346 \pm 0.0060$ & $16_{-11}^{+13}$ &  51 \\
                      & 0.5  &  463 & 1185 & $0.0369 \pm 0.0096$ & $13_{-11}^{+14}$ &  61 & $0.0262 \pm 0.0056$ & $14_{-11}^{+13}$ &  58 \\
\enddata
\tablecomments{Mode and mode-centric $1\sigma$ are presented for $\nbh$ for each GC. The `Max' $\nbh$ estimate corresponds to the $2\sigma$ upper bound for $\nbh$. Total mass in BHs can be easily estimated by assuming an average BH mass of $14\,\msun$, calculated in our models.
}
\label{T:results}
\end{deluxetable*}

%%%%%%%%%%%%%%%%%%%%%%%%%%%%%%%%%%%%%%%%%%%%%%%%%%%%%%%%%%%%%%%%%%%%%%%%%%%%%%%%%%%%%%%%%%%%%%%%%
%
\section{Summary} \label{S:summary}
%
%%%%%%%%%%%%%%%%%%%%%%%%%%%%%%%%%%%%%%%%%%%%%%%%%%%%%%%%%%%%%%%%%%%%%%%%%%%%%%%%%%%%%%%%%%%%%%%%%
We have presented the general method to indirectly estimate the number of stellar-mass BHs retained in GCs using measurements of mass segregation.
While the most reliable estimates would always come from specifically modeling individual clusters, that approach is computationally demanding. Instead, we used realistic models with a wide range in initial properties to determine a general relationship between measures of $\Delta$ and $\nbh/\ncluster$ (\autoref{S:models}). The number of BHs in any given GC can then be determined by measuring $\Delta$ (following our definitions) and $\ncluster$ (e.g., adopting a $M/L$ and average stellar mass) in the GC, making use of the model calibration that connects $\Delta$ to $\nbh/\ncluster$. 

We show examples of this process in full detail by estimating the retained $\nbh$ in three GCs (47\ Tuc, M\ 10, M\ 22) where candidate BHs have been reported, and where the required data exist as part of the ACS survey for MWGCs. Our analysis carefully takes into account observational constraints such as completeness and the radial extent of the existing data (\autoref{S:observational_results}). We find that among the three GCs investigated, M\ 22 contains the highest $\nbh/\ncluster$ (least mass-segregation), closely followed by M\ 10 and distantly trailed by 47 Tuc (highest mass-segregation; \autoref{S:nbh_predictions}; \autoref{f:9}). The relative measured values of $\Delta$ are generally consistent with past analysis of mass segregation in these GCs using very different methods \citep[e.g.,][]{2013ApJ...778...57G}. Furthermore, several past studies have shown that the larger the fraction of retained BHs, the larger the observed core radius \citep[e.g.,][]{2007MNRAS.379...93H,2017ApJ...834...68C,2018arXiv180205284A,2018arXiv180100795A}. This trend is reflected in our predicted results as well with M\ 22 having the largest and 47\ Tuc having the smallest $\rcobs$ among the three GCs (\autoref{T:props}). Interestingly, our predicted distribution of $\nbh$ in M\ 22 is consistent with the estimated $\nbh$ by \citet{2012Natur.490...71S} based on the two detected candidate BHs, although the latter estimate is based on specific assumptions of the donor type for the detected candidate BHs and expected duty cycles for X-ray binaries (\autoref{f:9}, \autoref{T:results}).  

Note that a central IMBH can also lead to a quenching of mass segregation \citep[e.g.,][]{2004ApJ...613.1143B,2007MNRAS.374..857T,2016ApJ...823..135P}. We did not take into account any IMBHs in our models. Theoretically, formation of IMBHs in dense star clusters has been well studied in the past \citep[e.g.,][]{2002ApJ...576..899P,2004Natur.428..724P,2006MNRAS.368..121F,2006MNRAS.368..141F,2010MNRAS.402..105G}. Their formation requires rather special conditions, such as primordial mass segregation and high initial concentration, which can lead to runaway stellar collisions before significant cluster expansion due to SN-driven mass loss \citep[e.g.,][]{2004ApJ...604..632G,2006MNRAS.368..141F,2012ApJ...752...43G}. If this indeed is the formation mechanism for IMBHs, then it is expected that almost all high-mass stars take part in the collisional runaway, leaving little chance to later produce significant numbers of stellar-mass BHs. 
Even if initial conditions suitable for the onset of a collisional runaway existed in some GC progenitors, it is unclear what actually happens to the massive collision product, i.e., does it grow through repeated collisions and followed by a direct collapse create a massive BH or does severe wind-mass loss limit growth \citep[e.g.,][]{2009arXiv0911.1483C,2009A&A...497..255G}. 
In any case, based on observed mass segregation, \citet{2016ApJ...823..135P} found no suitable candidate IMBH hosts among $\sim50$ of the MWGCs they analyzed. \citet{2017MNRAS.464.2174B} found that apart from $\omega$-Cen, no GCs in the MW show evidence of central IMBHs. Further constraints come from radio observations by \citet{2012ApJ...750L..27S}, who found no evidence of central IMBHs. 

Nevertheless, the dynamical effects of an IMBH and many centrally concentrated stellar-mass BHs of equivalent total mass would be quite similar and are likely indistinguishable using mass segregation alone \citep[e.g.,][]{2007MNRAS.379...93H}. Most recently, \citet{2017Natur.542..203K} suggested that 47\ Tuc may contain an IMBH of mass $2300_{-850}^{+1500}\,\msun$ (error bars are $1\sigma$) based on models that contained an IMBH but no stellar-mass BHs. Assuming $14\,\msun$ as the average BH mass in GCs (guided by our models) we find that the predicted total mass in stellar-mass BHs in 47\ Tuc can be up to $\sim2000\,\msun$ (up to $2\sigma$ confidence limit). Given the uncertainties, these numbers are not inconsistent with each other. Thus, the observed pulsar accelerations in 47\ Tuc may well be due to the presence of stellar-mass BHs rather than a central IMBH. However, detailed models dedicated for 47\ Tuc are necessary to make any strong predictions either way. 

In all of the GCs we have considered, the number of retained BHs is typically between $\sim$ a few tens to $\sim100$. This is likely common for most MWGCs with typical $\rcobs$ of up to a few $\pc$. Retained BH fractions are also strongly correlated with the observed core radius \citep[e.g.,][]{2007MNRAS.379...93H,2008MNRAS.386...65M,2017ApJ...834...68C,2018arXiv180100795A}. Cluster models that retain $\nbh$ in excess of $\sim10^3$ typically exhibit $\rcobs$ that are too large compared to the typical non-core-collapsed MWGCs. We have undertaken a wider survey of all MWGCs contained in the ACS survey catalog to estimate the number of BHs they retain at present. These determinations are expected to be extremely useful for the numerical modeling community since the final $\nbh/\ncluster$ will likely provide important constraints on the natal kick distributions for BHs which lacks either observational or theoretical constraints or somewhat equivalently, the dynamical ages of the GCs since the epoch of BH formation \citep{2018ApJ...855L..15K}. 

The intrinsic spread in $\nbh/\ncluster$ as a function of $\Delta$ (\autoref{f:3}, \autoref{f:9}) as well as the spread in the measured $\Delta$ stemming from the limited radial extent and completeness differences between populations (\autoref{f:9}) contribute to the wide spread in the predicted $\nbh/\ncluster$. The intrinsic spread typically comes from variations in metallicity, which leads to somewhat different BH mass spectrum, and binary fraction, which changes somewhat the dynamical energy production rate at the core and the average stellar mass of the cluster. The contribution from the uncertainty in the measured $\Delta$ can be improved via surveys that are more complete (e.g., \autoref{S:completeness}, \autoref{f:8}) even for fainter stars, and those that have large radial extents well above the $\rhl$. In any case, constraining $\nbh/\ncluster$ will remain hard at the limit where the number of retained BHs is dynamically unimportant. Nevertheless, we have demonstrated that using calibrations from realistic models and measuring $\Delta$ using existing data for the MW GCs, we can clearly distinguish GCs hosting a large population of BHs from those that host few. More stringent constraints can be achieved by targeted models of specific GCs \citep[e.g.,][]{2011MNRAS.410.2698G,2018ApJ...855L..15K} where all structural properties in addition to the measured $\Delta$ can be compared between models and observed clusters.  

%%%%%%%%%%%%%%%%%%%%%%%%%%%%%%%%%%%%%%%%%%%%%%%%%%%%%%%%%%%%%%%%%%%%%%%%%%%%%%%%%%%%%%%%%%%%%%%%%

\acknowledgements
We thank the anonymous referee for detailed and constructive comments. This work was made possible by Northwestern University (NU)'s High-Performance computing cluster \textit{Quest}, on which all CMC models were run. NW acknowledges support from the Illinois Space Grant Consortium, and a grant from NU for summer support. SC acknowledges Hubble Space Telescope Archival research grant HST-AR-14555.001-A (from the Space Telescope Science Institute, which is operated by the Association of Universities for Research in Astronomy, Incorporated, under NASA contract NAS5-26555). SC also acknowledges support from CIERA given as a fellowship. FAR acknowledges support from NASA ATP Grant NNX14AP92G and from NSF Grant AST-1716762.

\software
CMC \citep{2000ApJ...540..969J,2001ApJ...550..691J, 2003ApJ...593..772F, 2007ApJ...658.1047F, 2010ApJ...719..915C, 2012ApJ...750...31U, 2013MNRAS.429.2881C, 2013ApJS..204...15P, 2016MNRAS.463.2109R}. 

%\bibliography{bh_metallicity}

\begin{thebibliography}{}
\expandafter\ifx\csname natexlab\endcsname\relax\def\natexlab#1{#1}\fi
\providecommand{\url}[1]{\href{#1}{#1}}

\bibitem[{{Abbott} {et~al.}(2016{\natexlab{a}}){Abbott}, {Abbott}, {Abbott},
  {Abernathy}, {Acernese}, {Ackley}, {Adams}, {Adams}, {Addesso}, {Adhikari},
  \& et~al.}]{2016PhRvL.116f1102A}
{Abbott}, B.~P., {Abbott}, R., {Abbott}, T.~D., {et~al.} 2016{\natexlab{a}},
  Physical Review Letters, 116, 061102

\bibitem[{{Abbott} {et~al.}(2016{\natexlab{b}}){Abbott}, {Abbott}, {Abbott},
  {Abernathy}, {Acernese}, {Ackley}, {Adams}, {Adams}, {Addesso}, {Adhikari},
  \& et~al.}]{2016ApJ...818L..22A}
---. 2016{\natexlab{b}}, \apjl, 818, L22

\bibitem[{{Abbott} {et~al.}(2016{\natexlab{c}}){Abbott}, {Abbott}, {Abbott},
  {Abernathy}, {Acernese}, {Ackley}, {Adams}, {Adams}, {Addesso}, {Adhikari},
  \& et~al.}]{2016PhRvX...6d1015A}
---. 2016{\natexlab{c}}, Physical Review X, 6, 041015

\bibitem[{{Abbott} {et~al.}(2016{\natexlab{d}}){Abbott}, {Abbott}, {Abbott},
  {Abernathy}, {Acernese}, {Ackley}, {Adams}, {Adams}, {Addesso}, {Adhikari},
  \& et~al.}]{2016PhRvL.116x1103A}
---. 2016{\natexlab{d}}, Physical Review Letters, 116, 241103

\bibitem[{{Abbott} {et~al.}(2017){Abbott}, {Abbott}, {Abbott}, {Acernese},
  {Ackley}, {Adams}, {Adams}, {Addesso}, {Adhikari}, {Adya}, \&
  et~al.}]{2017PhRvL.118v1101A}
---. 2017, Physical Review Letters, 118, 221101

\bibitem[{{Alessandrini} {et~al.}(2016){Alessandrini}, {Lanzoni}, {Ferraro},
  {Miocchi}, \& {Vesperini}}]{2016ApJ...833..252A}
{Alessandrini}, E., {Lanzoni}, B., {Ferraro}, F.~R., {Miocchi}, P., \&
  {Vesperini}, E. 2016, \apj, 833, 252

\bibitem[{{Anderson} {et~al.}(2008){Anderson}, {Sarajedini}, {Bedin}, {King},
  {Piotto}, {Reid}, {Siegel}, {Majewski}, {Paust}, {Aparicio}, {Milone},
  {Chaboyer}, \& {Rosenberg}}]{2008AJ....135.2055A}
{Anderson}, J., {Sarajedini}, A., {Bedin}, L.~R., {et~al.} 2008, \aj, 135, 2055

\bibitem[{{Antonini} {et~al.}(2016){Antonini}, {Chatterjee}, {Rodriguez},
  {Morscher}, {Pattabiraman}, {Kalogera}, \& {Rasio}}]{2016ApJ...816...65A}
{Antonini}, F., {Chatterjee}, S., {Rodriguez}, C.~L., {et~al.} 2016, \apj, 816,
  65

\bibitem[{{Arca Sedda} {et~al.}(2018){Arca Sedda}, {Askar}, \&
  {Giersz}}]{2018arXiv180100795A}
{Arca Sedda}, M., {Askar}, A., \& {Giersz}, M. 2018, arXiv:1801.00795

\bibitem[{{Askar} {et~al.}(2018){Askar}, {Arca Sedda}, \&
  {Giersz}}]{2018arXiv180205284A}
{Askar}, A., {Arca Sedda}, M., \& {Giersz}, M. 2018, arXiv:1802.05284

\bibitem[{{Askar} {et~al.}(2017){Askar}, {Szkudlarek}, {Gondek-Rosi{\'n}ska},
  {Giersz}, \& {Bulik}}]{2017MNRAS.464L..36A}
{Askar}, A., {Szkudlarek}, M., {Gondek-Rosi{\'n}ska}, D., {Giersz}, M., \&
  {Bulik}, T. 2017, \mnras, 464, L36

\bibitem[{{Bahramian} {et~al.}(2017){Bahramian}, {Heinke}, {Tudor},
  {Miller-Jones}, {Bogdanov}, {Maccarone}, {Knigge}, {Sivakoff}, {Chomiuk},
  {Strader}, {Garcia}, \& {Kallman}}]{2017MNRAS.467.2199B}
{Bahramian}, A., {Heinke}, C.~O., {Tudor}, V., {et~al.} 2017, \mnras, 467, 2199

\bibitem[{{Banerjee}(2017)}]{2017MNRAS.467..524B}
{Banerjee}, S. 2017, \mnras, 467, 524

\bibitem[{{Banerjee}(2018{\natexlab{a}})}]{2018MNRAS.473..909B}
---. 2018{\natexlab{a}}, \mnras, 473, 909

\bibitem[{{Banerjee}(2018{\natexlab{b}})}]{2018arXiv180506466B}
---. 2018{\natexlab{b}}, arXiv:1805.06466

\bibitem[{{Banerjee} {et~al.}(2010){Banerjee}, {Baumgardt}, \&
  {Kroupa}}]{2010MNRAS.402..371B}
{Banerjee}, S., {Baumgardt}, H., \& {Kroupa}, P. 2010, \mnras, 402, 371

\bibitem[{{Baumgardt}(2017)}]{2017MNRAS.464.2174B}
{Baumgardt}, H. 2017, \mnras, 464, 2174

\bibitem[{{Baumgardt} {et~al.}(2004){Baumgardt}, {Makino}, \&
  {Ebisuzaki}}]{2004ApJ...613.1143B}
{Baumgardt}, H., {Makino}, J., \& {Ebisuzaki}, T. 2004, \apj, 613, 1143

\bibitem[{{Belczynski} {et~al.}(2010){Belczynski}, {Bulik}, {Fryer}, {Ruiter},
  {Valsecchi}, {Vink}, \& {Hurley}}]{2010ApJ...714.1217B}
{Belczynski}, K., {Bulik}, T., {Fryer}, C.~L., {et~al.} 2010, \apj, 714, 1217

\bibitem[{{Belczynski} {et~al.}(2002){Belczynski}, {Kalogera}, \&
  {Bulik}}]{2002ApJ...572..407B}
{Belczynski}, K., {Kalogera}, V., \& {Bulik}, T. 2002, \apj, 572, 407

\bibitem[{{Breen} \& {Heggie}(2013)}]{2013MNRAS.432.2779B}
{Breen}, P.~G., \& {Heggie}, D.~C. 2013, \mnras, 432, 2779

\bibitem[{{Chatterjee} {et~al.}(2010){Chatterjee}, {Fregeau}, {Umbreit}, \&
  {Rasio}}]{2010ApJ...719..915C}
{Chatterjee}, S., {Fregeau}, J.~M., {Umbreit}, S., \& {Rasio}, F.~A. 2010,
  \apj, 719, 915

\bibitem[{{Chatterjee} {et~al.}(2009){Chatterjee}, {Goswami}, {Umbreit},
  {Glebbeek}, {Rasio}, \& {Hurley}}]{2009arXiv0911.1483C}
{Chatterjee}, S., {Goswami}, S., {Umbreit}, S., {et~al.} 2009, arXiv:0911.1483

\bibitem[{{Chatterjee} {et~al.}(2013{\natexlab{a}}){Chatterjee}, {Rasio},
  {Sills}, \& {Glebbeek}}]{2013ApJ...777..106C}
{Chatterjee}, S., {Rasio}, F.~A., {Sills}, A., \& {Glebbeek}, E.
  2013{\natexlab{a}}, \apj, 777, 106

\bibitem[{{Chatterjee} {et~al.}(2017{\natexlab{a}}){Chatterjee}, {Rodriguez},
  {Kalogera}, \& {Rasio}}]{2017ApJ...836L..26C}
{Chatterjee}, S., {Rodriguez}, C.~L., {Kalogera}, V., \& {Rasio}, F.~A.
  2017{\natexlab{a}}, \apjl, 836, L26

\bibitem[{{Chatterjee} {et~al.}(2017{\natexlab{b}}){Chatterjee}, {Rodriguez},
  \& {Rasio}}]{2017ApJ...834...68C}
{Chatterjee}, S., {Rodriguez}, C.~L., \& {Rasio}, F.~A. 2017{\natexlab{b}},
  \apj, 834, 68

\bibitem[{{Chatterjee} {et~al.}(2013{\natexlab{b}}){Chatterjee}, {Umbreit},
  {Fregeau}, \& {Rasio}}]{2013MNRAS.429.2881C}
{Chatterjee}, S., {Umbreit}, S., {Fregeau}, J.~M., \& {Rasio}, F.~A.
  2013{\natexlab{b}}, \mnras, 429, 2881

\bibitem[{{Chen} \& {Han}(2009)}]{2009MNRAS.395.1822C}
{Chen}, X., \& {Han}, Z. 2009, \mnras, 395, 1822

\bibitem[{{Chomiuk} {et~al.}(2013){Chomiuk}, {Strader}, {Maccarone},
  {Miller-Jones}, {Heinke}, {Noyola}, {Seth}, \&
  {Ransom}}]{2013ApJ...777...69C}
{Chomiuk}, L., {Strader}, J., {Maccarone}, T.~J., {et~al.} 2013, \apj, 777, 69

\bibitem[{{Dalessandro} {et~al.}(2013){Dalessandro}, {Ferraro}, {Lanzoni},
  {Schiavon}, {O'Connell}, \& {Beccari}}]{2013ApJ...770...45D}
{Dalessandro}, E., {Ferraro}, F.~R., {Lanzoni}, B., {et~al.} 2013, \apj, 770,
  45

\bibitem[{{Fregeau} {et~al.}(2003){Fregeau}, {G{\"u}rkan}, {Joshi}, \&
  {Rasio}}]{2003ApJ...593..772F}
{Fregeau}, J.~M., {G{\"u}rkan}, M.~A., {Joshi}, K.~J., \& {Rasio}, F.~A. 2003,
  \apj, 593, 772

\bibitem[{{Fregeau} \& {Rasio}(2007)}]{2007ApJ...658.1047F}
{Fregeau}, J.~M., \& {Rasio}, F.~A. 2007, \apj, 658, 1047

\bibitem[{{Freitag} {et~al.}(2006{\natexlab{a}}){Freitag}, {G{\"u}rkan}, \&
  {Rasio}}]{2006MNRAS.368..141F}
{Freitag}, M., {G{\"u}rkan}, M.~A., \& {Rasio}, F.~A. 2006{\natexlab{a}},
  \mnras, 368, 141

\bibitem[{{Freitag} {et~al.}(2006{\natexlab{b}}){Freitag}, {Rasio}, \&
  {Baumgardt}}]{2006MNRAS.368..121F}
{Freitag}, M., {Rasio}, F.~A., \& {Baumgardt}, H. 2006{\natexlab{b}}, \mnras,
  368, 121

\bibitem[{{Fryer} {et~al.}(2012){Fryer}, {Belczynski}, {Wiktorowicz},
  {Dominik}, {Kalogera}, \& {Holz}}]{2012ApJ...749...91F}
{Fryer}, C.~L., {Belczynski}, K., {Wiktorowicz}, G., {et~al.} 2012, \apj, 749,
  91

\bibitem[{{Gaburov} {et~al.}(2010){Gaburov}, {Lombardi}, \& {Portegies
  Zwart}}]{2010MNRAS.402..105G}
{Gaburov}, E., {Lombardi}, Jr., J.~C., \& {Portegies Zwart}, S. 2010, \mnras,
  402, 105

\bibitem[{{Giersz} \& {Heggie}(2011)}]{2011MNRAS.410.2698G}
{Giersz}, M., \& {Heggie}, D.~C. 2011, \mnras, 410, 2698

\bibitem[{{Glebbeek} {et~al.}(2009){Glebbeek}, {Gaburov}, {de Mink}, {Pols}, \&
  {Portegies Zwart}}]{2009A&A...497..255G}
{Glebbeek}, E., {Gaburov}, E., {de Mink}, S.~E., {Pols}, O.~R., \& {Portegies
  Zwart}, S.~F. 2009, \aap, 497, 255

\bibitem[{{Goldsbury} {et~al.}(2013){Goldsbury}, {Heyl}, \&
  {Richer}}]{2013ApJ...778...57G}
{Goldsbury}, R., {Heyl}, J., \& {Richer}, H. 2013, \apj, 778, 57

\bibitem[{{Goswami} {et~al.}(2012){Goswami}, {Umbreit}, {Bierbaum}, \&
  {Rasio}}]{2012ApJ...752...43G}
{Goswami}, S., {Umbreit}, S., {Bierbaum}, M., \& {Rasio}, F.~A. 2012, \apj,
  752, 43

\bibitem[{{G{\"u}rkan} {et~al.}(2004){G{\"u}rkan}, {Freitag}, \&
  {Rasio}}]{2004ApJ...604..632G}
{G{\"u}rkan}, M.~A., {Freitag}, M., \& {Rasio}, F.~A. 2004, \apj, 604, 632

\bibitem[{{Hansen} \& {Kawaler}(1994)}]{1994sipp.book.....H}
{Hansen}, C.~J., \& {Kawaler}, S.~D. 1994, {Stellar Interiors. Physical
  Principles, Structure, and Evolution.}, 84

\bibitem[{{Harris}(1996)}]{1996AJ....112.1487H}
{Harris}, W.~E. 1996, \aj, 112, 1487

\bibitem[{{Heggie} \& {Hut}(2003)}]{2003gmbp.book.....H}
{Heggie}, D., \& {Hut}, P. 2003, {The Gravitational Million-Body Problem: A
  Multidisciplinary Approach to Star Cluster Dynamics}, Cambridge University Press

\bibitem[{{Heggie} \& {Giersz}(2014)}]{2014MNRAS.439.2459H}
{Heggie}, D.~C., \& {Giersz}, M. 2014, \mnras, 439, 2459

\bibitem[{{H{\'e}non}(1971{\natexlab{a}})}]{1971Ap&SS..13..284H}
{H{\'e}non}, M. 1971{\natexlab{a}}, \apss, 13, 284

\bibitem[{{H{\'e}non}(1971{\natexlab{b}})}]{1971Ap&SS..14..151H}
{H{\'e}non}, M.~H. 1971{\natexlab{b}}, \apss, 14, 151

\bibitem[{{Hurley}(2007)}]{2007MNRAS.379...93H}
{Hurley}, J.~R. 2007, \mnras, 379, 93

\bibitem[{{Hurley} {et~al.}(2000){Hurley}, {Pols}, \&
  {Tout}}]{2000MNRAS.315..543H}
{Hurley}, J.~R., {Pols}, O.~R., \& {Tout}, C.~A. 2000, \mnras, 315, 543

\bibitem[{{Hurley} {et~al.}(2016){Hurley}, {Sippel}, {Tout}, \&
  {Aarseth}}]{2016PASA...33...36H}
{Hurley}, J.~R., {Sippel}, A.~C., {Tout}, C.~A., \& {Aarseth}, S.~J. 2016,
  \pasa, 33, e036

\bibitem[{{Hurley} {et~al.}(2002){Hurley}, {Tout}, \&
  {Pols}}]{2002MNRAS.329..897H}
{Hurley}, J.~R., {Tout}, C.~A., \& {Pols}, O.~R. 2002, \mnras, 329, 897

\bibitem[{{Joshi} {et~al.}(2001){Joshi}, {Nave}, \&
  {Rasio}}]{2001ApJ...550..691J}
{Joshi}, K.~J., {Nave}, C.~P., \& {Rasio}, F.~A. 2001, \apj, 550, 691

\bibitem[{{Joshi} {et~al.}(2000){Joshi}, {Rasio}, \& {Portegies
  Zwart}}]{2000ApJ...540..969J}
{Joshi}, K.~J., {Rasio}, F.~A., \& {Portegies Zwart}, S. 2000, \apj, 540, 969

\bibitem[{{Kalogera} {et~al.}(2004){Kalogera}, {King}, \&
  {Rasio}}]{2004ApJ...601L.171K}
{Kalogera}, V., {King}, A.~R., \& {Rasio}, F.~A. 2004, \apjl, 601, L171

\bibitem[{{King}(1962)}]{1962AJ.....67..471K}
{King}, I. 1962, \aj, 67, 471

\bibitem[{{King}(1966)}]{1966AJ.....71...64K}
{King}, I.~R. 1966, \aj, 71, 64

\bibitem[{{King} {et~al.}(1995){King}, {Sosin}, \&
  {Cool}}]{1995ApJ...452L..33K}
{King}, I.~R., {Sosin}, C., \& {Cool}, A.~M. 1995, \apjl, 452, L33

\bibitem[{{K{\i}z{\i}ltan} {et~al.}(2017){K{\i}z{\i}ltan}, {Baumgardt}, \&
  {Loeb}}]{2017Natur.542..203K}
{K{\i}z{\i}ltan}, B., {Baumgardt}, H., \& {Loeb}, A. 2017, \nat, 542, 203

\bibitem[{{Kremer} {et~al.}(2018{\natexlab{a}}){Kremer}, {Chatterjee},
  {Rodriguez}, \& {Rasio}}]{2018ApJ...852...29K}
{Kremer}, K., {Chatterjee}, S., {Rodriguez}, C.~L., \& {Rasio}, F.~A.
  2018{\natexlab{a}}, \apj, 852, 29

\bibitem[{{Kremer} {et~al.}(2018{\natexlab{b}}){Kremer}, {Ye}, {Chatterjee},
  {Rodriguez}, \& {Rasio}}]{2018ApJ...855L..15K}
{Kremer}, K., {Ye}, C.~S., {Chatterjee}, S., {Rodriguez}, C.~L., \& {Rasio},
  F.~A. 2018{\natexlab{b}}, \apjl, 855, L15

\bibitem[{{Kroupa}(2001)}]{2001MNRAS.322..231K}
{Kroupa}, P. 2001, \mnras, 322, 231

\bibitem[{{Kulkarni} {et~al.}(1993){Kulkarni}, {Hut}, \&
  {McMillan}}]{1993Natur.364..421K}
{Kulkarni}, S.~R., {Hut}, P., \& {McMillan}, S. 1993, \nat, 364, 421

\bibitem[{{Larson}(1984)}]{1984MNRAS.210..763L}
{Larson}, R.~B. 1984, \mnras, 210, 763

\bibitem[{{Leigh} {et~al.}(2007){Leigh}, {Sills}, \&
  {Knigge}}]{2007ApJ...661..210L}
{Leigh}, N., {Sills}, A., \& {Knigge}, C. 2007, \apj, 661, 210

\bibitem[{{Li} {et~al.}(2018){Li}, {Deng}, {Bekki}, {Hong}, {de Grijs}, \&
  {For}}]{2018arXiv180707679L}
{Li}, C., {Deng}, L., {Bekki}, K., {et~al.} 2018, arXiv:1807.07679

\bibitem[{{Lombardi} {et~al.}(1996){Lombardi}, {Rasio}, \&
  {Shapiro}}]{1996ApJ...468..797L}
{Lombardi}, Jr., J.~C., {Rasio}, F.~A., \& {Shapiro}, S.~L. 1996, \apj, 468,
  797

\bibitem[{{Lombardi} {et~al.}(2002){Lombardi}, {Warren}, {Rasio}, {Sills}, \&
  {Warren}}]{2002ApJ...568..939L}
{Lombardi}, Jr., J.~C., {Warren}, J.~S., {Rasio}, F.~A., {Sills}, A., \&
  {Warren}, A.~R. 2002, \apj, 568, 939

\bibitem[{{Lombardi} {et~al.}(1995){Lombardi}, {Rasio}, \&
  {Shapiro}}]{1995ApJ...445L.117L}
{Lombardi}, C., J.~J., {Rasio}, F.~A., \& {Shapiro}, S.~L. 1995, \apjl, 445,
  L117

\bibitem[{{Maccarone} {et~al.}(2007){Maccarone}, {Kundu}, {Zepf}, \&
  {Rhode}}]{2007Natur.445..183M}
{Maccarone}, T.~J., {Kundu}, A., {Zepf}, S.~E., \& {Rhode}, K.~L. 2007, \nat,
  445, 183

\bibitem[{{Mackey} {et~al.}(2007){Mackey}, {Wilkinson}, {Davies}, \&
  {Gilmore}}]{2007MNRAS.379L..40M}
{Mackey}, A.~D., {Wilkinson}, M.~I., {Davies}, M.~B., \& {Gilmore}, G.~F. 2007,
  \mnras, 379, L40

\bibitem[{{Mackey} {et~al.}(2008){Mackey}, {Wilkinson}, {Davies}, \&
  {Gilmore}}]{2008MNRAS.386...65M}
---. 2008, \mnras, 386, 65

\bibitem[{{Mar{\'{\i}}n-Franch} {et~al.}(2009){Mar{\'{\i}}n-Franch},
  {Aparicio}, {Piotto}, {Rosenberg}, {Chaboyer}, {Sarajedini}, {Siegel},
  {Anderson}, {Bedin}, {Dotter}, {Hempel}, {King}, {Majewski}, {Milone},
  {Paust}, \& {Reid}}]{2009ApJ...694.1498M}
{Mar{\'{\i}}n-Franch}, A., {Aparicio}, A., {Piotto}, G., {et~al.} 2009, \apj,
  694, 1498

\bibitem[{{Marks} \& {Kroupa}(2010)}]{2010MNRAS.406.2000M}
{Marks}, M., \& {Kroupa}, P. 2010, \mnras, 406, 2000

\bibitem[{{Merritt} {et~al.}(2004){Merritt}, {Piatek}, {Portegies Zwart}, \&
  {Hemsendorf}}]{2004ApJ...608L..25M}
{Merritt}, D., {Piatek}, S., {Portegies Zwart}, S., \& {Hemsendorf}, M. 2004,
  \apjl, 608, L25

\bibitem[{{Miller-Jones} {et~al.}(2015){Miller-Jones}, {Strader}, {Heinke},
  {Maccarone}, {van den Berg}, {Knigge}, {Chomiuk}, {Noyola}, {Russell},
  {Seth}, \& {Sivakoff}}]{2015MNRAS.453.3918M}
{Miller-Jones}, J.~C.~A., {Strader}, J., {Heinke}, C.~O., {et~al.} 2015,
  \mnras, 453, 3918

\bibitem[{{Moody} \& {Sigurdsson}(2009)}]{2009ApJ...690.1370M}
{Moody}, K., \& {Sigurdsson}, S. 2009, \apj, 690, 1370

\bibitem[{{Morscher} {et~al.}(2015){Morscher}, {Pattabiraman}, {Rodriguez},
  {Rasio}, \& {Umbreit}}]{2015ApJ...800....9M}
{Morscher}, M., {Pattabiraman}, B., {Rodriguez}, C., {Rasio}, F.~A., \&
  {Umbreit}, S. 2015, \apj, 800, 9

\bibitem[{{Morscher} {et~al.}(2013){Morscher}, {Umbreit}, {Farr}, \&
  {Rasio}}]{2013ApJ...763L..15M}
{Morscher}, M., {Umbreit}, S., {Farr}, W.~M., \& {Rasio}, F.~A. 2013, \apjl,
  763, L15

\bibitem[{{Pasquato} {et~al.}(2016){Pasquato}, {Miocchi}, {Won}, \&
  {Lee}}]{2016ApJ...823..135P}
{Pasquato}, M., {Miocchi}, P., {Won}, S.~B., \& {Lee}, Y.-W. 2016, \apj, 823,
  135

\bibitem[{{Pattabiraman} {et~al.}(2013){Pattabiraman}, {Umbreit}, {Liao},
  {Choudhary}, {Kalogera}, {Memik}, \& {Rasio}}]{2013ApJS..204...15P}
{Pattabiraman}, B., {Umbreit}, S., {Liao}, W.-k., {et~al.} 2013, \apjs, 204, 15

\bibitem[{{Peuten} {et~al.}(2016){Peuten}, {Zocchi}, {Gieles}, {Gualandris}, \&
  {H{\'e}nault-Brunet}}]{2016MNRAS.462.2333P}
{Peuten}, M., {Zocchi}, A., {Gieles}, M., {Gualandris}, A., \&
  {H{\'e}nault-Brunet}, V. 2016, \mnras, 462, 2333

\bibitem[{{Portegies Zwart} {et~al.}(2004){Portegies Zwart}, {Baumgardt},
  {Hut}, {Makino}, \& {McMillan}}]{2004Natur.428..724P}
{Portegies Zwart}, S.~F., {Baumgardt}, H., {Hut}, P., {Makino}, J., \&
  {McMillan}, S.~L.~W. 2004, \nat, 428, 724

\bibitem[{{Portegies Zwart} \& {McMillan}(2000)}]{2000ApJ...528L..17P}
{Portegies Zwart}, S.~F., \& {McMillan}, S.~L.~W. 2000, \apjl, 528, L17

\bibitem[{{Portegies Zwart} \& {McMillan}(2002)}]{2002ApJ...576..899P}
---. 2002, \apj, 576, 899

\bibitem[{{Rodriguez} {et~al.}(2016{\natexlab{a}}){Rodriguez}, {Chatterjee}, \&
  {Rasio}}]{2016PhRvD..93h4029R}
{Rodriguez}, C.~L., {Chatterjee}, S., \& {Rasio}, F.~A. 2016{\natexlab{a}},
  \prd, 93, 084029

\bibitem[{{Rodriguez} {et~al.}(2016{\natexlab{b}}){Rodriguez}, {Haster},
  {Chatterjee}, {Kalogera}, \& {Rasio}}]{2016ApJ...824L...8R}
{Rodriguez}, C.~L., {Haster}, C.-J., {Chatterjee}, S., {Kalogera}, V., \&
  {Rasio}, F.~A. 2016{\natexlab{b}}, \apjl, 824, L8

\bibitem[{{Rodriguez} {et~al.}(2015){Rodriguez}, {Morscher}, {Pattabiraman},
  {Chatterjee}, {Haster}, \& {Rasio}}]{2015PhRvL.115e1101R}
{Rodriguez}, C.~L., {Morscher}, M., {Pattabiraman}, B., {et~al.} 2015, Physical
  Review Letters, 115, 051101

\bibitem[{{Rodriguez} {et~al.}(2016{\natexlab{c}}){Rodriguez}, {Morscher},
  {Wang}, {Chatterjee}, {Rasio}, \& {Spurzem}}]{2016MNRAS.463.2109R}
{Rodriguez}, C.~L., {Morscher}, M., {Wang}, L., {et~al.} 2016{\natexlab{c}},
  \mnras, 463, 2109

\bibitem[{{Sarajedini} {et~al.}(2007){Sarajedini}, {Bedin}, {Chaboyer},
  {Dotter}, {Siegel}, {Anderson}, {Aparicio}, {King}, {Majewski},
  {Mar{\'{\i}}n-Franch}, {Piotto}, {Reid}, \&
  {Rosenberg}}]{2007AJ....133.1658S}
{Sarajedini}, A., {Bedin}, L.~R., {Chaboyer}, B., {et~al.} 2007, \aj, 133, 1658

\bibitem[{{Scheepmaker} {et~al.}(2007){Scheepmaker}, {Haas}, {Gieles},
  {Bastian}, {Larsen}, \& {Lamers}}]{2007A&A...469..925S}
{Scheepmaker}, R.~A., {Haas}, M.~R., {Gieles}, M., {et~al.} 2007, \aap, 469,
  925

\bibitem[{{Sigurdsson} \& {Hernquist}(1993)}]{1993Natur.364..423S}
{Sigurdsson}, S., \& {Hernquist}, L. 1993, \nat, 364, 423

\bibitem[{{Sills} {et~al.}(2001){Sills}, {Faber}, {Lombardi}, {Rasio}, \&
  {Warren}}]{2001ApJ...548..323S}
{Sills}, A., {Faber}, J.~A., {Lombardi}, Jr., J.~C., {Rasio}, F.~A., \&
  {Warren}, A.~R. 2001, \apj, 548, 323

\bibitem[{{Sills} {et~al.}(2013){Sills}, {Glebbeek}, {Chatterjee}, \&
  {Rasio}}]{2013ApJ...777..105S}
{Sills}, A., {Glebbeek}, E., {Chatterjee}, S., \& {Rasio}, F.~A. 2013, \apj,
  777, 105

\bibitem[{{Sills} {et~al.}(1997){Sills}, {Lombardi}, {Bailyn}, {Demarque},
  {Rasio}, \& {Shapiro}}]{1997ApJ...487..290S}
{Sills}, A., {Lombardi}, Jr., J.~C., {Bailyn}, C.~D., {et~al.} 1997, \apj, 487,
  290

\bibitem[{{Sirianni} {et~al.}(2005){Sirianni}, {Jee}, {Ben{\'{\i}}tez},
  {Blakeslee}, {Martel}, {Meurer}, {Clampin}, {De Marchi}, {Ford}, {Gilliland},
  {Hartig}, {Illingworth}, {Mack}, \& {McCann}}]{2005PASP..117.1049S}
{Sirianni}, M., {Jee}, M.~J., {Ben{\'{\i}}tez}, N., {et~al.} 2005, \pasp, 117,
  1049

\bibitem[{{Spitzer}(1969)}]{1969ApJ...158L.139S}
{Spitzer}, Jr., L. 1969, \apjl, 158, L139

\bibitem[{{Strader}(2014)}]{2014cxo..prop.4353S}
{Strader}, J. 2014, {A Black Hole in the Galactic Globular Cluster M10},
  Chandra Proposal, ,

\bibitem[{{Strader} {et~al.}(2012{\natexlab{a}}){Strader}, {Chomiuk},
  {Maccarone}, {Miller-Jones}, \& {Seth}}]{2012Natur.490...71S}
{Strader}, J., {Chomiuk}, L., {Maccarone}, T.~J., {Miller-Jones}, J.~C.~A., \&
  {Seth}, A.~C. 2012{\natexlab{a}}, \nat, 490, 71

\bibitem[{{Strader} {et~al.}(2012{\natexlab{b}}){Strader}, {Chomiuk},
  {Maccarone}, {Miller-Jones}, {Seth}, {Heinke}, \&
  {Sivakoff}}]{2012ApJ...750L..27S}
{Strader}, J., {Chomiuk}, L., {Maccarone}, T.~J., {et~al.} 2012{\natexlab{b}},
  \apjl, 750, L27

\bibitem[{{Trenti} {et~al.}(2007){Trenti}, {Ardi}, {Mineshige}, \&
  {Hut}}]{2007MNRAS.374..857T}
{Trenti}, M., {Ardi}, E., {Mineshige}, S., \& {Hut}, P. 2007, \mnras, 374, 857

\bibitem[{{Umbreit}(2012)}]{2012Natur.490...46U}
{Umbreit}, S. 2012, \nat, 490, 46

\bibitem[{{Umbreit} {et~al.}(2012){Umbreit}, {Fregeau}, {Chatterjee}, \&
  {Rasio}}]{2012ApJ...750...31U}
{Umbreit}, S., {Fregeau}, J.~M., {Chatterjee}, S., \& {Rasio}, F.~A. 2012,
  \apj, 750, 31

\bibitem[{{Vink} {et~al.}(2001){Vink}, {de Koter}, \&
  {Lamers}}]{2001A&A...369..574V}
{Vink}, J.~S., {de Koter}, A., \& {Lamers}, H.~J.~G.~L.~M. 2001, \aap, 369, 574

\bibitem[{{Wang} {et~al.}(2016){Wang}, {Spurzem}, {Aarseth}, {Giersz}, {Askar},
  {Berczik}, {Naab}, {Schadow}, \& {Kouwenhoven}}]{2016MNRAS.458.1450W}
{Wang}, L., {Spurzem}, R., {Aarseth}, S., {et~al.} 2016, \mnras, 458, 1450

\bibitem[{{Webb} \& {Vesperini}(2016)}]{2016MNRAS.463.2383W}
{Webb}, J.~J., \& {Vesperini}, E. 2016, \mnras, 463, 2383

\bibitem[{{Webb} {et~al.}(2017){Webb}, {Vesperini}, {Dalessandro}, {Beccari},
  {Ferraro}, \& {Lanzoni}}]{2017MNRAS.471.3845W}
{Webb}, J.~J., {Vesperini}, E., {Dalessandro}, E., {et~al.} 2017, \mnras, 471,
  3845

\bibitem[{{Ziosi} {et~al.}(2014){Ziosi}, {Mapelli}, {Branchesi}, \&
  {Tormen}}]{2014MNRAS.441.3703Z}
{Ziosi}, B.~M., {Mapelli}, M., {Branchesi}, M., \& {Tormen}, G. 2014, \mnras,
  441, 3703

\end{thebibliography}

\end{document}